\documentclass[prd,aps,nofootinbib]{revtex4}
\usepackage{amssymb,amsmath,epsfig,bm}
\usepackage[bookmarks]{hyperref}
\usepackage[usenames,dvipsnames]{color}

\newcommand{\Real}{\mathbb{R}}
\newcommand{\Complex}{\mathbb{C}}
\newcommand{\re}{\mbox{Re}}
\newcommand{\im}{\mbox{Im}}

\newcommand{\gtens}{\mbox{\boldmath $g$}}
\newcommand{\Atens}{\mbox{\boldmath $A$}}
\newcommand{\Ftens}{\mbox{\boldmath $F$}}
\newcommand{\gtiltens}{\tilde{\gtens}}

\newcommand{\gtil}{\tilde{g}}

\newcommand{\laptil}{\tilde{\Delta}}
\newcommand{\nabtil}{\tilde{\nabla}}

\newcommand{\varepsilontiltens}{\tilde{\bm\varepsilon}}

\newtheorem{lemma}{Lemma}

\newcommand{\proof}{\noindent {\bf Proof. }}

\newcommand{\qed}{\hfill $\fbox{\hspace{0.3mm}}$ \vspace{.3cm}} 

\begin{document}

\title{Black holes in nonlinear electrodynamics: quasi-normal spectra and parity splitting}

\author{Eliana Chaverra$^{1,2,3}$, Juan Carlos Degollado$^4$, Claudia Moreno$^5$, and Olivier Sarbach$^1$}
\affiliation{$^1$Instituto de F\'isica y Matem\'aticas,
Universidad Michoacana de San Nicol\'as de Hidalgo,\\
Edificio C-3, Ciudad Universitaria, 58040 Morelia, Michoac\'an, M\'exico,\\
$^2$Escuela Nacional de Estudios Superiores, Unidad Morelia, Universidad Nacional Aut\'onoma de M\'exico, Campus Morelia, C.P. 58190, Morelia, Michoac\'an, M\'exico,\\
$^3$Departamento de F\'isica, Divisi\'on de Ciencias e Ingenier\'ia, Campus Le\'on, 
Universidad de Guanajuato, Le\'on 37150, M\'exico,\\
$^4$Instituto de Ciencias F\'isicas, Universidad Nacional Aut\'onoma de M\'exico,
Apdo. Postal 48-3, 62251, Cuernavaca, Morelos, M\'exico.\\
$^5$Departamento de F\'isica,
Centro Universitario de Ciencias Exactas e Ingenier\'ias, Universidad de Guadalajara, 
Avenida Revoluci\'on 1500, Colonia Ol\'impica C.P. 44430, Guadalajara, Jalisco, M\'exico.}

\begin{abstract}
We discuss the quasi-normal oscillations of black holes which are sourced by a nonlinear electrodynamic field. 
While previous studies have focused on the computation of quasi-normal frequencies for the wave or higher spin 
equation on a fixed background geometry described by such black holes, here we compute for the first time the quasi-normal 
frequencies for the coupled electromagnetic-gravitational linear perturbations.

To this purpose, we consider a parametrized family of Lagrangians for the electromagnetic field which contains the 
Maxwell Lagrangian as a special case. In the Maxwell case, the unique spherically symmetric black hole solutions 
are described by the Reissner-Nordstr\"om family and in this case it is well-known that the quasi-normal spectra 
in the even- and odd-parity sectors are identical to each other. However, when moving away from the Maxwell case, 
we obtain deformed Reissner-Nordstr\"om black holes, and we show that in this case there is a parity splitting in 
the quasi-normal mode spectra. A partial explanation for this phenomena is provided by considering the eikonal 
(high-frequency) limit.
\end{abstract}

\maketitle

\section{Introduction}
\label{Sect:Intro}

Black holes are undoubtedly among the most fascinating objects in general relativity and at the forefront of modern astrophysical observations~\cite{LIGOPRL16,sDetal08}. Despite the complexity of Einstein's field equations, asymptotically flat, stationary black holes usually arise with a very simple structure and are uniquely characterized by a small set of global parameters such as their mass, angular momentum and charges, much like the elementary particles in high-energy physics. This fact is due to the no-hair theorems which are valid in pure vacuum, electrovacuum (general relativity coupled to a Maxwell field) and to certain general classes of theories involving scalar fields coupled to Abelian gauge fields (see~\cite{pCjCmH12} for a recent review). Furthermore, an inherent property of most of the physically relevant black hole solutions is that they contain a curvature singularity in their interior.

There has been quite some interest in coupling general relativity to non-standard electromagnetic theories, described by Lagrangians which are a nonlinear function of the invariants $F := \frac{1}{4} F^{\mu\nu} F_{\mu\nu}$ and $\tilde{F} := \frac{1}{4} F^{\mu\nu} * F_{\mu\nu}$ built from the Faraday tensor ${\bf F} = \frac{1}{2} F_{\mu\nu} dx^\mu\wedge dx^\nu$ and its Hodge dual $\ast {\bf F}$. In fact, such Lagrangians were introduced a long time ago by Born and Infeld~\cite{mBlI34} who constructed a theory in which static field configurations with finite energy exist and are interpreted as describing an elementary particle of matter such as the electron. More recently, it has been shown that nonlinear electromagnetic fields may be coupled to gravity to yield singularity-free black hole solutions~\cite{eAaG98,eAaG99a,eAaG99b,eAaG00,eAaG05}, as well as globally regular soliton-like solutions~\cite{kB01}. This can even be achieved by requiring the validity of the weak energy condition and by demanding that the nonlinear electromagnetic Lagrangians have the correct weak field limit, so that the corresponding theories reduce to the usual Einstein-Maxwell theory for weak electromagnetic fields. Apart from their own theoretical interest and importance in understanding the necessary hypothesis underlying the singularity theorems (see~\cite{jB68} 
and~\cite{HawkingEllis-Book} and references therein), recently, singularity-free black holes have also been used to model black hole candidates in astrophysical observations~\cite{Bambi:2014nta,Stuchlik:2014qja,Li:2013jra}. Further recent developments in nonlinear electrodynamics (NED) include the study of light propagation in the geometric optics approximation (see for instance~\cite{yOgR02,sF12,gSvPcL16}), the formulation of well-posed Cauchy problems in these theories~\cite{fAfCeGoR15}, and the description of the dark sector of the Universe to explain the accelerated expansion phase~\cite{Elizalde:2003ku,Novello:2006ng,Breton:2012yt}.

In this article we focus on the dynamical response of black holes in NED to small external perturbations. We are particularly interested in the question of whether or not this response exhibits qualitatively different features than perturbed black holes in standard electrodynamics.

When a black hole is perturbed by a sufficiently small perturbation, one usually finds that it relaxes to an equilibrium state during which (gravitational, electromagnetic or other, depending on the underlying theory) radiation is emitted. The characteristic features of this radiation can be described in three stages. The first stage corresponds to a burst of radiation which depends strongly on the details of the initial conditions describing the perturbation. However, this burst is followed by a second stage corresponding to a ringdown phase, during which the emitted signal as well as the black hole oscillate with well-defined (complex) frequencies. These frequencies, called quasi-normal (QN) frequencies, do not depend on the initial conditions and are only dependent on the properties of the final equilibrium black hole. During the third and last stage, all multipoles of the fields decay according to a power law (the so-called tail decay).

The study of QN black hole spectra is a broad an active topic in general relativity, see~\cite{hN99,kKbS99,eBvCaS09} for reviews. From the observational point of view, a ringdown signal has already been observed in the recent first detection of gravitational waves from the coalescence of two black holes~\cite{LIGOPRL16}, and it is expected that QN frequencies will soon be measured in a systematic way by the gravitational wave antennas such as LIGO~\cite{LIGO}, VIRGO~\cite{VIRGO} and KAGRA~\cite{KAGRA}. Since the QN spectra only depend on the geometric properties of the black hole, which in turn are characterized by a small set of parameters, the detection and measurement of QN frequencies provides a valuable tool which allows not only to determine the parameters characterizing the black hole such as its mass and spin, but also to test the underlying theory. From the theoretical point of view, the AdS/CFT duality has lead to an increase on studying QN modes of black holes with a negative cosmological constant, see for instance~\cite{Cardoso:2013pza,Cardoso:2003cj,cW15} and references therein. 

In this work we compute for the first time the fundamental QN frequencies associated with the gravitational and electrodynamic perturbations of black holes in NED theories. There has been much previous work discussing QN modes for regular black holes in NED, see for instance~\cite{sF05,sFcH06,sFjC12,aFjL13,bTaAzSbA15,jLkLnY15,sF15,pXxA16} where comparisons with the QN frequencies of the Reissner-Nordstr\"om (RN) black holes were also carried out by varying the charge, mass and the spherical harmonic index. However, we would like to stress that these results only apply to \emph{test fields} propagating on the (fixed) geometry of such black holes. In contrast to this, in the present work we take into account the (linearized) perturbations of the metric and the electromagnetic field in a self-consistent way. We base our calculations on the pulsation master equations derived some time ago by two of us~\cite{cMoS03}, which have the form of wave equations with a matrix-valued potential which describes the coupling between the gravitational and electromagnetic perturbations.

As has been shown by Moncrief~\cite{vM74c, vM74d,vM75}, in the RN case of linear electrodynamics the master equations can be decoupled into two scalar equations in each parity sector. Furthermore, in the RN case, there exist intertwining relations which connects the even- and odd-parity potentials~\cite{Chandrasekhar-Book} and imply that the QN modes are \emph{isospectral}, that is, identical in both parity sectors. However, for black holes in NED it does not seem to be possible to decouple the master equations in an explicit way, and hence the question of whether or not the even- and odd-parity potentials can be related to each other via an intertwining relation becomes cumbersome. In fact, we show in this article that for deformed RN black holes in NED theories the QN spectrum ceases to be isospectral. This parity splitting phenomena can be partially understood by considering the eikonal limit, in which the QN frequencies with high values of the angular momentum number $\ell$ can be related to the properties of the unstable, circular null geodesics. However, in contrast to the standard Maxwell theory in which the light rays in the geometric optics approximation are described by null geodesics of the gravitational metric ${\bf g}$ of spacetime, in NED the light rays are null geodesics belonging to either one of two effective metrics $\mathfrak{g}_1$ or $\mathfrak{g}_2$ which depend on both the spacetime metric ${\bf g}$ and the Faraday tensor ${\bf F}$. As we show, the fact that $\mathfrak{g}_1\neq \mathfrak{g}_2$ is directly related to the parity splitting phenomena, at least in the eikonal limit.

The remainder of this article is structured as follows. In Sec.~\ref{Sect:EM} we review the equations of motion governing a self-gravitating, nonlinear electromagnetic field configuration and discuss spherically symmetric, purely magnetic black hole solutions in such theories. The linearized field equations around such background solutions have been analyzed in detail in Ref.~\cite{cMoS03}, and in Sec.~\ref{Sect-Puls} we first review these perturbation equations. As indicated above, they can be cast into a family of wave equations which are coupled through an effective $2\times 2$-matrix-valued potential whose structure is also studied in Sec.~\ref{Sect-Puls}. Further, the particular case of RN perturbations is discussed, and it is shown how the wave equations can be decoupled in this case, following early work by Moncrief. Next, in Sec.~\ref{Sec-Num} we perform numerical Cauchy evolutions of the perturbation equations, using horizon-penetrating coordinates, and show that typical initial pulses for the perturbations give rise to QN oscillations whose frequencies as observed by static observers are measured. A semi-analytic direct method for computing the corresponding QN frequencies and modes is discussed in Sec.~\ref{Sec-QNM}. This method is a straightforward generalization to systems of wave equations of the method first described in Ref.~\cite{eCmMoS15}, where applications to the oscillations of accretion flows are analyzed (see also~\cite{eCoS16}). The eikonal limit and tests for the RN case are also discussed in Sec.~\ref{Sec-QNM}. In particular, we show that our method is able to reproduce accurately known results in the literature for the RN case. Next, in Sec.~\ref{Sec:Results} we compute the QN frequencies for a certain class of deformed RN black holes which are obtained from a specific, parametrized nonlinear Lagrangian ${\cal L}_\eta(F)$ of the electromagnetic invariant $F = \frac{1}{4} F^{\mu\nu} F_{\mu\nu}$. The case $\eta=0$ corresponds to the RN case, and here we recover the known results in which the QN spectrum is identical in both parity sectors. However, when $\eta > 0$, corresponding to the deformed RN black holes, we find that this symmetry between the parity sectors is broken. In Sec.~\ref{Sec:Results}, we also compute the QN frequencies for values of the angular momentum number $\ell$ between $2$ and $9$ and compare them to the results obtained in the eikonal limit based on the two effective metrics $\mathfrak{g}_1$ and $\mathfrak{g}_2$, showing good qualitative agreement and providing a partial explanation for the parity splitting phenomena. Finally, conclusions of our work are drawn in Sec.~\ref{Sec:Conclusions} and a detailed analysis of the deformed RN black holes studied in this article is given in an appendix.

Our signature convention for the metric is $(-,+,+,+)$ with Greek letters denoting spacetime indices. We choose units for which Newton's constant and the speed of light are one.

\section{Equations of motion}
\label{Sect:EM}

The action describing the dynamics of a self-gravitating nonlinear electromagnetic field in general relativity is
\begin{equation}
{\cal S}\left[ \gtens,\Atens \right] = \frac{1}{4\pi}\int\left[ \frac{1}{4} R - {\cal L}(F) \right] 
\sqrt{-g}\, d^4 x ,
\end{equation}
where here $R$ denotes the Ricci scalar with respect to the spacetime metric $\gtens$, ${\cal L}$ is a smooth function of the invariant $F\equiv \frac{1}{4} F^{\mu\nu} F_{\mu\nu}$ with $\Ftens = \frac{1}{2} F_{\mu\nu} dx^\mu\wedge dx^\nu = d\Atens$ the Faraday tensor, and where $g$ denotes the determinant of the metric. In Einstein-Maxwell theory, ${\cal L}(F) = F$, but here we consider more general choices of electromagnetic Lagrangians.\footnote{However, in this article we do not consider the case where ${\cal L}$ depends on the invariant $\tilde{F} = \frac{1}{4} F_{\mu\nu}\ast F^{\mu\nu}$ as well.} The stationary points of the action functional ${\cal S}$ which are obtained by setting its first variation with respect to compactly supported perturbations yield the following field equations:
\begin{eqnarray}
&& G^\mu{}_\nu = 2\left[ {\cal L}_F F^{\mu\alpha} F_{\nu\alpha} 
- \delta^\mu{}_\nu{\cal L} \right],
\label{Eq:EinsteinF}\\
&& d\left( {\cal L}_F\ast\Ftens \right) = 0,
\label{Eq:NMaxF}\\
&& d\Ftens = 0,
\label{Eq:NBianchiF}
\end{eqnarray}
where $G^\mu{}_\nu$ denote the components of the Einstein tensor, ${\cal L}_F \equiv \partial {\cal L}/\partial F$, and $\ast$ refers to the Hodge dual.\footnote{In local coordinates, the Hodge dual $\ast\Ftens$ of the Faraday tensor is given by $\ast F^{\mu\nu} = \frac{1}{2} \varepsilon^{\mu\nu\alpha\beta} F_{\alpha\beta}$, where $\varepsilon_{\mu\nu\alpha\beta}$ is completely skew-symmetric with $\varepsilon_{0123} = \sqrt{-g}$.} Here, for completeness and later use, we have included Eq.~(\ref{Eq:NBianchiF}) which follows directly from $\Ftens = d\Atens$.

We shall require the weak energy condition which states that the stress energy-momentum tensor ${\bf T}$ satisfies $T_{\mu\nu} u^\mu u^\nu \geq 0$ for all timelike vector fields ${\bf u}$, meaning that all causal observers measure a nonnegative energy density. By virtue of the Einstein field equations~(\ref{Eq:EinsteinF}) we see that this is equivalent to the condition
\begin{equation}
{\cal L}_F E^\alpha E_\alpha + {\cal L} \geq 0,
\label{Eq:WEC}
\end{equation}
with $E_\alpha := u^\mu F_{\mu\alpha}$ the electric field measured by an observer with four-velocity ${\bf u}$.

\subsection{Spherically symmetric, purely magnetic solutions}

A spherically symmetric spacetime is the product of a two-dimensional pseudo-Riemannian manifold $(\tilde{M},\gtiltens)$ with the unit two sphere $(S^2,d\Omega^2)$,
\begin{displaymath}
M = \tilde{M} \times S^2,\qquad
\gtens = \gtil_{ab}\, dx^a dx^b + r^2 d\Omega^2, 
\end{displaymath}
where $\gtiltens = \gtil_{ab}\, dx^a dx^b$ is a metric of signature $(-1,1)$ on the manifold $\tilde{M}$, $r$ is a positive function on $\tilde{M}$, describing the areal radius of the invariant $2$-spheres, and $d\Omega^2 = d\vartheta^2 + \sin^2\vartheta\, d\varphi^2$ is the standard metric on the two-sphere $S^2$.

Purely magnetic solutions are obtained from the following ansatz for the Faraday tensor:
\begin{equation}
\Ftens = Q_m d\Omega = Q_m d\vartheta\wedge \sin\vartheta d\varphi,
\label{Eq:Faraday}
\end{equation}
with $Q_m$ the (constant) magnetic charge. This implies that
\begin{equation}
\ast\Ftens = \frac{Q_m}{r^2}\varepsilontiltens,\qquad
F = \frac{Q_m^2}{2r^4},
\end{equation}
with $\varepsilontiltens$ the natural volume form on $(\tilde{M},\gtiltens)$, and one verifies easily that the nonlinear Maxwell equations~(\ref{Eq:NMaxF},\ref{Eq:NBianchiF}) are automatically satisfied. With this ansatz for the electromagnetic field, Einstein's field equations~(\ref{Eq:EinsteinF}) reduce to
\begin{eqnarray*}
G^a{}_b &=& -2\delta^a{}_b {\cal L},\\
G^A{}_b &=& 0,\\
G^A{}_B &=& 2\delta^A{}_B{\cal H},\qquad
{\cal H} := 2{\cal L}_F F - {\cal L},
\end{eqnarray*}
where $A,B = \vartheta,\varphi$. In order to develop this further we choose local coordinates $(t,r)$ on $\tilde{M}$ which include the areal radius $r$ and are such that the two-metric $\gtiltens$ is diagonal,
\begin{equation}
\gtiltens = -S^2 N dt^{\, 2} +\frac{dr^2}{N},\qquad N := 1 - \frac{2m}{r},
\label{Eq:SchwCoord}
\end{equation}
with $S$ and $m$ functions of $(t,r)$. Here, $m$ is the Misner-Sharp mass~\cite{cMdS64} which is invariantly defined through the equation $\gtiltens(dr,dr) = N = 1 - 2m/r$. With this parametrization for the two-metric, the radial components of the Einstein tensor are
\begin{equation}
G^t{}_t = -\frac{2m'}{r^2},\qquad
G^r{}_r = \frac{2NS'}{rS} - \frac{2m'}{r^2},\qquad
G^r{}_t = \frac{2\dot{m}}{r^2},
\end{equation}
where a prime and a dot denote partial differentiation with respect to $r$ and $t$, respectively. Therefore, it follows from Einstein's field equations that $S = S(t)$ (which can be set to $1$ by an appropriate redefinition of $t$), that $m$ is time-independent 
and that $m' = r^2{\cal L}$. Therefore, the metric is static for $r > 2m(r)$ and is given by
\begin{equation}
\gtens = -\left( 1 - \frac{2m(r)}{r} \right) dt^2 + \frac{dr^2}{1 - \frac{2m(r)}{r}}
 + r^2 d\Omega^2,\qquad
m(r) = M 
 - \int\limits_r^\infty \bar{r}^2 {\cal L}\left( F=\frac{Q_m^2}{2\bar{r}^4} \right)d\bar{r},
\label{Eq:MetricSol}
\end{equation}
with $M$ the ADM mass of the configuration. For linear electrodynamics, $m(r) = M - Q_m^2/(2r)$ and one recovers from Eqs.~(\ref{Eq:Faraday},\ref{Eq:MetricSol}) the magnetically charged RN solution. For Lagrangians ${\cal L}$ satisfying the weak field limit, ${\cal L}(F) \simeq F$ for small $F$, this solution is still valid asymptotically, as $r\to\infty$, and in this case the metric has RN asymptotics.

The solutions we are interested in here correspond to choices for the Lagrangians ${\cal L}$ which are finite in the strong field regime, $F\to \infty$. More specifically, we shall assume that for large $F$, ${\cal L}(F) = \Lambda/2 + {\cal O}(1/\sqrt{F})$ with $\Lambda$ a positive constant. This implies that the integral in Eq.~(\ref{Eq:MetricSol}) converges for $r\to 0$, and we obtain the asymptotic behavior
\begin{equation}
m(r) = m_0 + \frac{\Lambda}{6} r^3 + {\cal O}(r^5),\qquad
m_0 := M
 - \int\limits_0^\infty \bar{r}^2 {\cal L}\left( F=\frac{Q_m^2}{2\bar{r}^4} \right)d\bar{r}, 
\label{Eq:mrto0}
\end{equation}
for small $r$. Therefore, when $m_0\neq 0$ the behavior close to $r=0$ is similar to the one of the Schwarzschild-de-Sitter solution. The particular case when $m_0=0$, that is, when the integral exactly balances the ADM mass, gives rise to a solution which is regular at $r=0$ since in this case $2m(r)/r = {\cal O}(r^2)$.

\subsection{The Bardeen black holes}

As has been shown in Ref.~\cite{eAaG00} the family of Bardeen black holes~\cite{jB68} can be obtained from the Lagrangian
\begin{equation}
{\cal L}(y) = \frac{3}{2sg^2}\left( \frac{y}{1+y} \right)^{5/2},
\label{Eq:BardeenModel}
\end{equation}
where $y = \sqrt{2g^2F}$ and $s$ and $g$ are two positive constants such that $g^2 F$ is dimensionless. The first three derivates of ${\cal L}$ with respect to $y$ are
\begin{eqnarray}
{\cal L}_y &=& \frac{15}{4sg^2}\frac{ y^{3/2} }{(1+y)^{7/2}},
\label{Eq:LBy}\\
{\cal L}_{yy} &=& \frac{15}{8sg^2} \frac{ y^{1/2} }{(1+y)^{9/2}} (3-4y),
\label{Eq:LByy}\\ 
{\cal L}_{yyy} &=& \frac{45}{16sg^2} \frac{ y^{-1/2} }{(1+y)^{11/2}} (8y^2-12y+1),
\label{Eq:LByyy}
\end{eqnarray}
and the fact that ${\cal L}$ and ${\cal L}_y$ are both positive implies the validity of the weak energy condition, see Eq.~(\ref{Eq:WEC}).
The family of Lagrangians~(\ref{Eq:BardeenModel}) has the required strong field limit, ${\cal L}(F) = 3/(2sg^2) + {\cal O}(1/\sqrt{F})$ for $F\to\infty$; notice however that is does not have the correct weak field limit.

Nevertheless, the integral in Eq.~(\ref{Eq:MetricSol}) converges, and one obtains the expression
\begin{equation}
m(r) = m_0 + (M - m_0)\frac{r^3}{(\alpha^2 + r^2)^{3/2}},\qquad
m_0 := M - \frac{|\alpha|^3}{2sg^2},
\label{Eq:BardeenSol}
\end{equation}
for the mass function, where here and in the following we introduce the quantity
\begin{equation}
\alpha := \sqrt{\frac{g}{|Q_m|}} Q_m.
\label{Eq:alphaDef}
\end{equation}
For $r\to 0$ the function $m(r)$ defined in Eq.~(\ref{Eq:BardeenSol}) has the asymptotic form~(\ref{Eq:mrto0}) while for $r\to\infty$ we obtain $m(r) = M - 3|\alpha|^5[ 1 + {\cal O}(\alpha^2/r^2)]/(4sg^2 r^2)$, showing that the spacetime is asymptotically flat with ADM mass $M$. For generic values of the free parameters $Q_m$ and $M$, the spacetime metric corresponding to (\ref{Eq:BardeenSol}) is singular at $r=0$. However, in case the magnetic charge $Q_m$ and the ADM mass $M$ are related to  each other via $|Q_m|^3 = 4s^2g M^2$, the constant $m_0$ vanishes and one obtains a solution which is regular at the center $r=0$:
\begin{equation}
N(r) = 1 - \frac{2m(r)}{r} = 1 - \frac{2M r^2}{(r^2 + \alpha^2)^{3/2}},\qquad
\alpha^2 = g|Q_m|.
\label{Eq:BardeenBH}
\end{equation}
The function $N(r)$ has a global minimum at $r = r_{min} := \sqrt{2}|\alpha|$, where $N(r_{min}) = 1 - 4M/(3^{3/2}|\alpha|)$. 
Therefore, the metric~(\ref{Eq:MetricSol}) represents a regular black hole which has an event and a Cauchy horizon when $|\alpha|/M < 4/3^{3/2}$, and when $|\alpha|/M > 4/3^{3/2}$ it represents a soliton.

In the literature, one often sets $|Q_m| = g$ and $s = g/(2M)$ in which case the condition for regularity is automatically satisfied. However, here we shall adopt the point of view where we fix the theory first, that is the Lagrangian ${\cal L}$ and its parameters $(s,g)$, and then consider all its spherically symmetric black hole solutions which are parametrized by $(Q_m,M)$. From this point of view, clearly, the choice $|Q_m|^3 = 4s^2 g M^2$ leading to the regular solutions is special.

\subsection{An alternative model}
\label{SubSec:Alternative}

An alternative to the Bardeen black hole which does have a correct weak field limit is obtained from the Lagrangian\footnote{For other alternative models in the electric sector, see~\cite{eAaG98,eAaG99a,eAaG99b,eAaG05}.}
\begin{equation}
{\cal L}(y) = \frac{1}{2g^2}\left( \frac{y}{1+\gamma y} \right)^2,\qquad
y = \sqrt{2g^2 F},
\label{Eq:AlternativeModel}
\end{equation}
with constants $g > 0$ and $\gamma\geq 0$. When $\gamma = 0$ this reduces to the Maxwell case. In general, when $\gamma y\ll 1$, ${\cal L}(y) \simeq y^2/(2g^2) = F$, while for $\gamma y \gg 1$,
$$
{\cal L}(y) = \frac{1}{2g^2\gamma^2}\left[ 1 - \frac{2}{\gamma y} 
 + {\cal O}\left(\frac{1}{\gamma y} \right)^2 \right],
$$
so that this Lagrangian has the required strong and weak field limits. The first three derivatives of ${\cal L}(y)$ are
\begin{eqnarray}
{\cal L}_y &=& \frac{1}{g^2}\frac{y}{(1+\gamma y)^3},
\label{Eq:LAy}\\
{\cal L}_{yy} &=& \frac{1}{g^2} \frac{1-2\gamma y}{(1+\gamma y)^4},
\label{Eq:LAyy}\\ 
{\cal L}_{yyy} &=& \frac{1}{g^2} \frac{6\gamma(\gamma y - 1)}{(1+\gamma y)^5},
\label{Eq:LAyyy}
\end{eqnarray}
and the fact that ${\cal L}$ and ${\cal L}_y$ are both positive implies the validity of the weak energy condition, see Eq.~(\ref{Eq:WEC}).
Using the relation $y = g|Q_m|/r^2$ the solution for the metric function $N(r)$ is obtained from  Eq.~(\ref{Eq:MetricSol}) and yields
\begin{equation}
N(r) = 1 - \frac{2M}{r} + \frac{Q_m^2}{2}\left[ 
 \frac{1}{r^2 + \eta^2} + \frac{\arctan\left( \frac{\eta}{r} \right)}{\eta r} \right].
\label{Eq:AlternativeBH}
\end{equation}
with $\eta := \sqrt{\gamma g |Q_m|}$.

As was expected from the validity of the correct weak field limit, the metric behaves asymptotically as the RN metric when $r\to \infty$; that is
$$
N(r) = 1 - \frac{2M}{r} + \frac{Q_m^2}{r^2} + {\cal O}\left( \frac{1}{r^4} \right).
$$
The metric is singular at $r = 0$ unless
$$
M = \frac{\pi}{8}\frac{Q_m^2}{\eta}.
$$
The global behavior of the function $N(r)$ as a function of the parameters $\eta$, $M$ and $Q_m$ is analyzed in Appendix~\ref{App:SolutionProp}. Provided $Q_m^2/(M\eta)$ is small enough, the metric always describes a singular black holes. For higher values, the solution either describes a singular black hole, a regular black hole, a regular soliton or a naked singularity, see Appendix~\ref{App:SolutionProp} for details. In the following we will focus on the (singular or regular) black hole case.

\section{Perturbation equations}
\label{Sect-Puls}

In this section we start with a brief review of the results presented in Ref.~\cite{cMoS03} regarding the master equations describing even- and odd-parity electro-gravitational linear perturbations of the spherically symmetric, magnetic black holes described in the previous section. These equations have the form of a system of two wave equations which are coupled to each other through a matrix-valued, symmetric potential. The structure of the potential is then analyzed in more detail and explicit expressions for the alternative model are given.

\subsection{Summary of previous results}

The pulsation equations, describing the propagation of linearized gravitational and electromagnetic perturbations about the soliton and black hole solutions described in the previous section, can be obtained from the equations in Ref.~\cite{cMoS03} after applying the $FP$ duality described in Eq.~(8) of that reference. The inverse transformation implies
\begin{equation}
P\mapsto -F,\qquad
{\cal H}\mapsto -{\cal L},\qquad
{\cal H}_P\mapsto {\cal L}_F,\qquad
{\cal H}_{PP}\mapsto -{\cal L}_{FF}.
\end{equation}
The resulting pulsation equations for $\ell\geq 2$ are a coupled wave system of the form
\begin{equation}
\tilde{\Box}\left( \begin{array}{l} \Psi_{\ell m\pm} \\ \Phi_{\ell m\pm} \end{array} \right)
 + V_{\ell\pm}(r)
 \left( \begin{array}{l} \Psi_{\ell m\pm} \\ \Phi_{\ell m\pm} \end{array} \right) = 0,
\label{Eq:Pulsation}
\end{equation}
where $\tilde{\Box} := -\gtil^{ab}\nabtil_a\nabtil_b$ is the covariant d'Alembertian on $(\tilde{M},\gtiltens)$, $\Psi_{\ell m\pm}$ and $\Phi_{\ell m\pm}$ are gauge-invariant perturbation amplitudes, describing the linearized gravitational and electromagnetic fluctuations with angular momentum numbers $\ell m$ and parity $\pm$, and $V_{\ell\pm}(r)$ is a matrix-valued potential which is symmetric.

In the odd-parity sector ($-$), the potential is
\begin{equation}
V_{\ell-}(r) = \left( \begin{array}{ll}
\frac{\ell(\ell + 1)}{r^2} - \frac{6m}{r^3} + 2{\cal L} 
 & -\frac{ \sqrt{4\lambda {\cal L}_F}\, Q_m}{r^3} \\
-\frac{ \sqrt{4\lambda {\cal L}_F}\, Q_m}{r^3}
 & \frac{\ell(\ell+1)}{r^2} + {\cal L}_F^{1/2}\laptil {\cal L}_F^{-1/2} 
 + \frac{4 Q_m^2}{r^4} {\cal L}_F 
\end{array} \right),
\end{equation}
where here, $\laptil := \gtil^{ab}\nabtil_a\nabtil_b = \partial_r(N\partial_r)$ is the covariant Laplacian on $(\tilde{M},\gtiltens)$ and $\lambda := (\ell-1)(\ell+2)$. We have also used the background equation
\begin{equation}
r\laptil r = \frac{2m}{r} - 2r^2{\cal L},
\end{equation}
in deriving this potential from the results in~\cite{cMoS03}.

In the even-parity sector the potential is
\begin{equation}
V_{\ell+}(r) = \left( \begin{array}{ll}
V_{11}(r) & -\frac{ \sqrt{4\lambda {\cal L}_F}\, Q_m}{r^3} W(r) \\
-\frac{ \sqrt{4\lambda {\cal L}_F}\, Q_m}{r^3} W(r) & V_{22}(r)
\end{array} \right),
\end{equation}
where the functions $V_{11}$, $V_{22}$ and $W$ are defined as
\begin{eqnarray*}
V_{11}(r) &=& \frac{1}{r^2(a + \lambda)}\left[ \ell(\ell+1)\lambda -2N\lambda 
 + a\left( a - \frac{4m}{r} \right) \right] + \frac{2N\lambda\, b}{r^2(a + \lambda)^2},\\
V_{22}(r) &=& \kappa\frac{\ell(\ell+1)}{r^2}
 + \frac{ 4{\cal L}_F Q_m^2}{r^4(a + \lambda)}\left( \lambda + 1 - N + 2r^2 {\cal L} 
 + 4N\kappa \right) + {\cal L}_F^{-1/2}\laptil {\cal L}_F^{1/2}
  + \frac{8N{\cal L}_F Q_m^2\, b}{r^4(a + \lambda)^2},\\
W(r) &=& \frac{1}{a + \lambda}\left( \lambda + 1 - N + 2r^2 {\cal L} + 2N\kappa \right) + \frac{2N\, b}{(a + \lambda)^2},
\end{eqnarray*}
with
\begin{equation*}
a := \frac{6m}{r} - 2r^2 {\cal L},\qquad
b := \lambda + 4{\cal L}_F\frac{Q_m^2}{r^2},\qquad
\kappa := y {\cal L}_y^{-1} {\cal L}_{yy} = 1 + 2F {\cal L}_F^{-1} {\cal L}_{FF}.
\end{equation*}
The results presented so far are valid for an arbitrary choice of the Lagrangian ${\cal L}(F)$ satisfying the condition ${\cal L}_F > 0$.

\subsection{Linear stability criteria}
\label{SubSec:Stability}

In~\cite{cMoS03} conditions on the Lagrangian ${\cal L}$ and the metric function $N$ were derived which guarantee the corresponding black hole solutions to be stable under linear perturbations. These conditions are the satisfaction of the inequalities
\begin{equation}
{\cal L} > 0,\qquad
{\cal L}_y > 0,\qquad
{\cal L}_{yy} > 0,\qquad
N\kappa \leq 3,
\label{Eq:StabilityCond}
\end{equation}
outside the event horizon. 
Furthermore, it was shown in~\cite{cMoS03} that if $\kappa$ is negative in an (arbitrarily small open) region outside the event horizon, the black hole is unstable with respect to linear, even-parity perturbations with high enough angular momentum $\ell$.

One can show~\cite{cMoS03} that the sufficient conditions~(\ref{Eq:StabilityCond}) are always satisfied for the Bardeen black holes described in Eq.~(\ref{Eq:BardeenBH}) with $g = |Q_m|$ and $s = g/(2M)$. The satisfaction of these conditions for the deformed RN black holes in the alternative model are discussed in Appendix~\ref {App:SolutionProp}.

\subsection{The structure of the effective potential}
\label{SubSec:Structure}

For the analysis below, it is convenient to consider ${\cal L}$ to be a function of the dimensionless variable $y := \sqrt{2g^2 F}$ instead of $F$, as in the two model Lagrangians~(\ref{Eq:BardeenModel},\ref{Eq:AlternativeModel}) discussed above. For the spherically symmetric, purely magnetic ansatz~(\ref{Eq:Faraday}) it follows that $y = \alpha^2/r^2$ with $\alpha$ defined as in Eq.~(\ref {Eq:alphaDef}). In terms of the first three partial derivatives ${\cal L}_y$, ${\cal L}_{yy}$, ${\cal L}_{yyy}$ of the Lagrangian with respect to $y$ one finds the following expressions:
\begin{eqnarray*}
a &=& \frac{6m}{r} - 2r^2 {\cal L},\\
b &=& \lambda + 4\alpha^2{\cal L}_y,\\
\kappa &=& y{\cal L}_y^{-1}{\cal L}_{yy},\\
{\cal L}_F^{1/2}\laptil {\cal L}_F^{-1/2} &=& 
 \frac{1}{r^2}\left( a - \frac{4m}{r} \right)(\kappa-1) + \frac{N}{r^2} h_-(y),\\
{\cal L}_F^{-1/2}\laptil {\cal L}_F^{1/2} &=& 
-\frac{1}{r^2}\left( a - \frac{4m}{r} \right)(\kappa-1) - \frac{N}{r^2} h_+(y),
\end{eqnarray*}
with
\begin{eqnarray}
h_-(y) &:=& 2 - 5y{\cal L}_y^{-1}{\cal L}_{yy} + 3y^2{\cal L}_y^{-2}{\cal L}_{yy}^2
 - 2y^2{\cal L}_y^{-1} {\cal L}_{yyy},
\label{Eq:p-}\\
h_+(y) &:=& -y{\cal L}_y^{-1}{\cal L}_{yy} + y^2{\cal L}_y^{-2}{\cal L}_{yy}^2
 - 2y^2{\cal L}_y^{-1} {\cal L}_{yyy}.
\label{Eq:p+}
\end{eqnarray}
Using this, one can rewrite the effective potentials in the form
\begin{equation}
V_{\ell-}(r) = \frac{\ell(\ell+1)}{r^2}\left( \begin{array}{ll} 1 & 0 \\ 0 & 1 \end{array} \right)
 + \frac{1}{r^2}\left( \begin{array}{cc} 
 -a & -\sqrt{4\alpha^2\lambda{\cal L}_y} \\ 
 -\sqrt{4\alpha^2\lambda{\cal L}_y}  & \left( a - \frac{4m}{r} \right)(\kappa-1) + N h_-(y)
 + 4\alpha^2{\cal L}_y
 \end{array} \right),
\label{Eq:EffPotOdd}
\end{equation}
in the odd-parity sector. In the even-parity sector, we obtain
\begin{equation}
V_{\ell+}(r) = \frac{\ell(\ell+1)}{r^2}
\left( \begin{array}{ll} 1 & 0 \\ 0 & \kappa \end{array} \right)
 + \frac{1}{r^2}\left( \begin{array}{cc} 
 -a + b_{11}(r) & -\sqrt{4\alpha^2\lambda{\cal L}_y}[1 + w(r)] \\ 
 -\sqrt{4\alpha^2\lambda{\cal L}_y}[1 + w(r)]  &  
 -\left( a - \frac{4m}{r} \right)(\kappa-1) - N h_+(y) + 4\alpha^2{\cal L}_y + b_{22}(r)
\end{array} \right),
\label{Eq:EffPotEven}
\end{equation}
with
\begin{eqnarray}
b_{11}(r) &=& \frac{2}{a + \lambda}\left[ a(a-2) + 4N\alpha^2 {\cal L}_y \right]
 + \frac{2Na}{(a+\lambda)^2}\left[ a - 4\alpha^2{\cal L}_y \right],\\
b_{22}(r) &=& \frac{8\alpha^2{\cal L}_y}{a + \lambda}
\left[ 2 - a + N(2\kappa-1) \right]
 - \frac{8N\alpha^2{\cal L}_y}{(a+\lambda)^2}\left[ a - 4\alpha^2{\cal L}_y \right],\\
w(r) &=& \frac{2}{a + \lambda}\left[ 2-a + N(\kappa-1) \right]
 - \frac{2N}{(a+\lambda)^2} \left[ a - 4\alpha^2{\cal L}_y \right].
\end{eqnarray}

Here, we recall that $a = 6m/r - 2r^2{\cal L}$, $\kappa = y{\cal L}_y^{-1}{\cal L}_{yy}$ and that the functions $h_\pm(y)$, which depend on the first three partial derivatives of ${\cal L}$ with respect to the dimensionless variable $y = \alpha^2/r^2$, are defined in Eqs.~(\ref{Eq:p-},\ref{Eq:p+}). In deriving this result, we have expanded the even-parity potential in inverse powers of $a + \lambda = a + (\ell-1)(\ell+2)$. From this form of writing the effective potential one can make the following interesting observations:
\begin{enumerate}
\item[(i)] For linear electromagnetism, where ${\cal L} = y^2/(2g^2)$, one obtains $a = 6M/r - 4Q_m^2/r^2$, $\alpha\sqrt{4{\cal L}_y} = 2Q_m/r$, $\kappa=1$ and $h_\pm(y) = 0$, and the potentials reduce to the ones obtained by Moncrief~\cite{vM74c, vM74d,vM75}. In this case, the equations can be decoupled from each other by a suitable constant linear transformation, and the problem can be reduced to the analysis of purely scalar wave equations, see below.
\item[(ii)] When $\alpha=0$ and ${\cal L}=0$, the equations for $\Psi_{\ell m\pm}$ decouple, and they reduce to the well-known Regge-Wheeler~\cite{tRjW57} and Zerilli~\cite{fZ70} equations.
\item[(iii)] For a Lagrangian ${\cal L}$ satisfying the weak-field limit, that is, ${\cal L}$ is proportional to $y^2$ for small $y$, it follows that $\kappa\to 1$, $h_\pm(y)\to 0$, and the potential is dominated by the usual centrifugal term when $r\to\infty$. Interestingly though, in the Bardeen model one has $\kappa\to 3/2$ instead which yields a non-trivial factor in front of the centrifugal term in the even-parity potential.
\item[(iv)] In the limit $\ell\to\infty$ the functions $b_{11}(r)$, $b_{22}(r)$, $w(r)$ vanish and in this case the two potentials $V_{\ell-}$ and $V_{\ell+}$ are quite similar to each other, the only differences relying in the factor $\kappa$ in the centrifugal term and the different signs in the lower right entry. For linear electromagnetism, where $\kappa=1$ and $h_\pm(y) = 0$, these differences go away. Therefore, the different behavior of the potential for high angular resolutions is entirely due to nonlinearities in the model. As we will see later, these differences in the two parity sectors have consequences for the QN spectra.
\item[(v)] The aforementioned linear instability which occurs when $\kappa < 0$ can be understood by noticing that for high values of $\ell$ the second diagonal element of $V_{\ell+}$ is dominated by $\ell(\ell+1)\kappa/r^2$, which is negative in the region where $\kappa < 0$, and hence leads to an exponentially growing mode. See~\cite{cMoS03} for a rigorous proof of this statement.
\end{enumerate}

In terms of the standard Schwarzschild-like coordinates $(t,r)$ defined in Eq.~(\ref{Eq:MetricSol}) the perturbation equations~(\ref{Eq:Pulsation}) can be rewritten as
\begin{equation}
\left[ \frac{\partial^2}{\partial t^2} -  \frac{\partial^2}{\partial r_*^2} 
 + N(r) V_{\ell\pm}(r) \right] 
 \left( \begin{array}{l} \Psi_{\ell m\pm} \\ \Phi_{\ell m\pm} \end{array} \right) = 0,
\label{Eq:PulsationBis}
\end{equation}
where $N(r) = 1 - 2m(r)/r$ and the tortoise coordinate $r_*$ is related to $r$ via the formula
\begin{equation}
r_* = \int\limits^r \frac{dr'}{N(r')}.
\label{Eq:Tortoise}
\end{equation}

\subsection{Decoupling in the RN case}
\label{SubSec:DecouplingRN}

Since it turns out to be relevant for the interpretation of the QN modes discussed in the next sections, here we first discuss the particular case of linear electromagnetism. As mentioned above, in this case it is possible to decouple the $2\times 2$ wave system~(\ref{Eq:PulsationBis}). To see why this is the case we first notice that (in the general case) we can rewrite the effective potentials $V_{\ell\pm}$ in the form
\begin{equation}
V_{\ell-}(r) = \frac{\ell(\ell+1)}{r^2}\left( \begin{array}{ll} 1 & 0 \\ 0 & 1 \end{array} \right)
 + \frac{1}{r^2}\left( \begin{array}{cc} 
 -A_- & 0 \\
 0 & -A_- + \left( a - \frac{4m}{r} \right)(\kappa-1) + N h_-(y)
 \end{array} \right) + \frac{1}{r^3} T,
\label{Eq:EffPotRN-}
\end{equation}
and
\begin{eqnarray}
V_{\ell+}(r) &=& \frac{\ell(\ell+1)}{r^2}\left( \begin{array}{ll} 1 & 0 \\ 0 & \kappa \end{array} \right)
 + \frac{1}{r^2}\left( \begin{array}{cc}
C - A_-\left(1 - \frac{2(\kappa-1)N}{a + \lambda} \right) & 0 \\
 0 & C - A_-\left(1 + \frac{2(\kappa-1)N}{a + \lambda} \right)
 - \left( a - \frac{4m}{r} \right)(\kappa-1) - N h_+(y)
 \end{array} \right)\nonumber\\
 &+& \frac{1+w}{r^3} T,
\label{Eq:EffPotRN+}
\end{eqnarray}
with $A_\pm := (a \pm 4\alpha^2{\cal L}_y)/2$,
$$
C := \frac{1}{2}(b_{22} + b_{11})
 = \frac{2}{a + \lambda}\left[ (a-2)A_- + 4\kappa\alpha^2{\cal L}_y N \right]
 + \frac{4N A_-^2}{(a+\lambda)^2},
$$
and where the matrix $T$ is given by
$$
T := r\left( \begin{array}{cc} 
 -A_+ & -\sqrt{4\alpha^2\lambda{\cal L}_y} \\
 -\sqrt{4\alpha^2\lambda{\cal L}_y} & A_+
 \end{array} \right).
$$
In the RN case $\kappa=1$, $h_\pm(y) = 0$, and so the first two terms in the above expressions for $V_{\ell\pm}$ are proportional to the identity matrix. Furthermore, in this case we have $A_+ = 3M/r$ and $4\alpha^2{\cal L}_y = 4Q_m^2/r^2$, such that $T$ becomes the constant matrix
$$
T = \left( \begin{array}{cc} 
 -3M & -\sqrt{4\lambda Q_m^2} \\
 -\sqrt{4\lambda Q_m^2} & +3M
 \end{array} \right),
$$
which has the eigenvalues $\mu_1 = \sqrt{9M^2 + 4\lambda Q_m^2}$, $\mu_2 = -\sqrt{9M^2 + 4\lambda Q_m^2}$ and the corresponding orthogonal constant eigenvectors
$$
e_1 := \left( \begin{array}{c}
 -\sqrt{4\lambda Q_m^2} \\
 3M + \sqrt{9M^2 + 4\lambda Q_m^2}
\end{array} \right),\qquad
e_2 := \left( \begin{array}{c}
 -\sqrt{4\lambda Q_m^2} \\
 3M - \sqrt{9M^2 + 4\lambda Q_m^2}
\end{array} \right).
$$
Therefore, in the RN case, the wave system can be decoupled by means of the transformation
\begin{equation}
 \left( \begin{array}{l} \Psi_{\ell m\pm}(t,r) \\ \Phi_{\ell m\pm}(t,r) \end{array} \right)
 = Z_1^{(\pm)}(t,r) e_1 + Z_2^{(\pm)}(t,r) e_2,
 \label{eq:transf_z}
\end{equation}
where for notational simplicity we have dropped the indices $\ell m$ in the definitions of the decoupled wave functions $Z_{1,2}^{(\pm)}$. 
Then, each function $Z_i^{(\pm)}$ satisfies a \emph{scalar} wave equation of the form
$$
\left[ \frac{\partial^2}{\partial t^2} -  \frac{\partial^2}{\partial r_*^2} 
 + N(r) W_i^{(\pm)}(r) \right] Z_i^{(\pm)} = 0,\qquad i = 1,2,
$$
where the explicit expressions for $W_i^{(\pm)}(r)$ which will not be needed here can be found in Refs.~\cite{vM74c, vM74d,vM75, Chandrasekhar-Book}. As in the Schwarzschild case it can be shown that (for fixed $i$) the two potentials $W_i^{(-)}$ and $W_i^{(+)}$ are connected to each other via an intertwining relation which implies, in particular, that the QN modes are isospectral~\cite{Chandrasekhar-Book}.

However, in NED it does not seem possible to decouple the $2\times 2$ wave system~(\ref{Eq:PulsationBis}) in this fashion. Therefore, in this case one has to deal with the coupled system of equations. A method for dealing with such coupled systems will be discussed in Sec.~\ref{Sec-QNM}.

\subsection{Explicit expressions for the alternative model}
\label{SubSec:Explicit}

The results given so far are valid for any spherically symmetric, purely magnetic background solution. From now on, for definiteness, we focus our attention on the model introduced in Sec.~\ref{SubSec:Alternative}. In this case, we find the following explicit expressions:
\begin{eqnarray*}
m(r) &=& M - \frac{Q_m^2}{4}\left[
 \frac{r}{r^2 + \eta^2} + \frac{\arctan\left( \frac{\eta}{r} \right)}{\eta} \right],\\
2r^2 {\cal L} &=& Q_m^2\frac{r^2}{(r^2 + \eta^2)^2},\\
\sqrt{4\alpha^2{\cal L}_y} &=& \frac{2|Q_m|}{r}\left( 1 + \frac{\eta^2}{r^2} \right)^{-3/2},\\
\kappa &=& \frac{1 - 2\frac{\eta^2}{r^2}}{1 + \frac{\eta^2}{r^2}},\\
h_-(y) &=& \frac{\eta^2}{r^2}\frac{9 + 12\frac{\eta^2}{r^2}}{\left( 1 + \frac{\eta^2}{r^2} \right)^2},\\
h_+(y) &=& \frac{\eta^2}{r^2}\frac{9 - 6\frac{\eta^2}{r^2}}{\left( 1 + \frac{\eta^2}{r^2} \right)^2}.
\end{eqnarray*}
The effective potentials $V_{\ell\pm}(r)$ given by Eqs.~(\ref{Eq:EffPotOdd},\ref{Eq:EffPotEven}) can be explicitly computed from these expressions.

\section{Numerical evolution of the pulsation equations}
\label{Sec-Num}

In this section we solve the pulsation equations in the time domain, and compute numerically the signal originating from an initial perturbation as measured by a static observer. In order to do so, we use ingoing Eddington-Finkelstein-type coordinates $(\tilde t,r,\vartheta,\varphi)$, obtained from the Schwarzschild-like coordinates~(\ref{Eq:MetricSol}) by transforming the time coordinate as $\tilde t = t +(r_*-r)$, with $r_*$ the tortoise coordinate defined in Eq.~(\ref{Eq:Tortoise}). In terms of the new coordinates the metric becomes
\begin{equation}
{\bf g} = - (\tilde\alpha^2-\beta^{r}\beta_{r}) d\tilde t^{2} + 2\beta_{r}drd\tilde t + \gamma_{rr}dr^2 +r^2 d\Omega^2,
\end{equation}
with the lapse and radial component of the shift vector given by
\begin{equation}
\tilde\alpha = \frac{1}{\sqrt{2-N(r)}} = \frac{1}{\sqrt{1 + \frac{2m(r)}{r}}}\ , \quad 
\beta^{r} = \frac{1-N(r)}{2-N(r)}= \tilde\alpha^{2}\beta_{r} \ , \quad \beta_{r} = \frac{2m(r)}{r}.
 \end{equation}
Writing the metric in this form has become popular in numerical simulations because a hypersurface of constant $\tilde t$ is non-singular at the horizon and thus the inner boundary of the computational domain can be placed inside the black hole.

For simplicity, let us denote the gauge-invariant perturbations $\Psi_{\ell m\pm}$ and $\Phi_{\ell m\pm}$ by $R^{(1)}_{\pm}$ and $R^{(2)}_{\pm}$ in the following. To reformulate the pulsation equations as a first-order system of equations, we introduce the auxiliary fields
\begin{equation}
\psi^{(i)}_{\pm} := \partial_{r}R^{(i)}_{\pm} \ ,  \qquad {\rm and } \qquad
\pi^{(i)}_{\pm} := \frac{1}{\tilde\alpha^2} \left( \partial_{\tilde t}R^{(i)}_{\pm} -\beta^{r}\psi^{(i)}_{\pm}  \right) \ .
 \end{equation}
With these new variables the equations of motion can be written as
\begin{eqnarray}
\label{eq:evolution}
\partial_{\tilde t} R^{(i)}_{\pm} &=& \tilde\alpha^2 \pi^{(i)}_{\pm} + \beta^{r}\psi^{(i)}_{\pm} \ ,\\ \nonumber
 \partial_{\tilde t} \psi^{(i)}_{\pm} &=& \partial_{r}(\tilde\alpha^2 \pi^{(i)}_{\pm} + \beta^{r}\psi^{(i)}_{\pm}) \ , \\ \nonumber
\tilde\alpha^2 \partial_{\tilde t} \pi^{(i)}_{\pm} &=&  -\beta^{r}\tilde\alpha^2\partial_{r}\pi^{(i)}_{\pm} -[(\beta^r)^2-N^2 ]\partial_{r}\psi^{(i)}_{\pm}-
 \beta^{r}[\pi^{(i)}_{\pm}\partial_{r}\tilde\alpha^2+\psi^{(i)}_{\pm} \partial_{r}\beta^{r}]+NN'\psi^{(i)}_{\pm} -N (V_{\ell\pm}^{(i,1)}R^{(1)}_{\pm}+V_{\ell\pm}^{(i, 2)}R^{(2)}_{\pm}),
 \end{eqnarray}
where $V_{\ell\pm}^{(i,j)}$ denote the components of the matrix-valued potentials $V_{\ell\pm}$ defined in Eqs.~(\ref{Eq:EffPotOdd},\ref{Eq:EffPotEven}).

We solved the evolution system~(\ref{eq:evolution}) by making use of a $1+1$ dimensional PDE code as 
described in~\cite{Degollado:2009rw,Nunez:2010ra}. The time evolution was based on the method of lines
with a third order Runge-Kutta scheme. The spatial derivatives were evaluated using a second-order symmetric 
finite difference stencil. In order to suppress potential high-frequency instabilities a standard fourth order 
dissipation term was also applied.

\subsection{Results}

\begin{figure}[h!]
\centering
\includegraphics[height=2.8in]{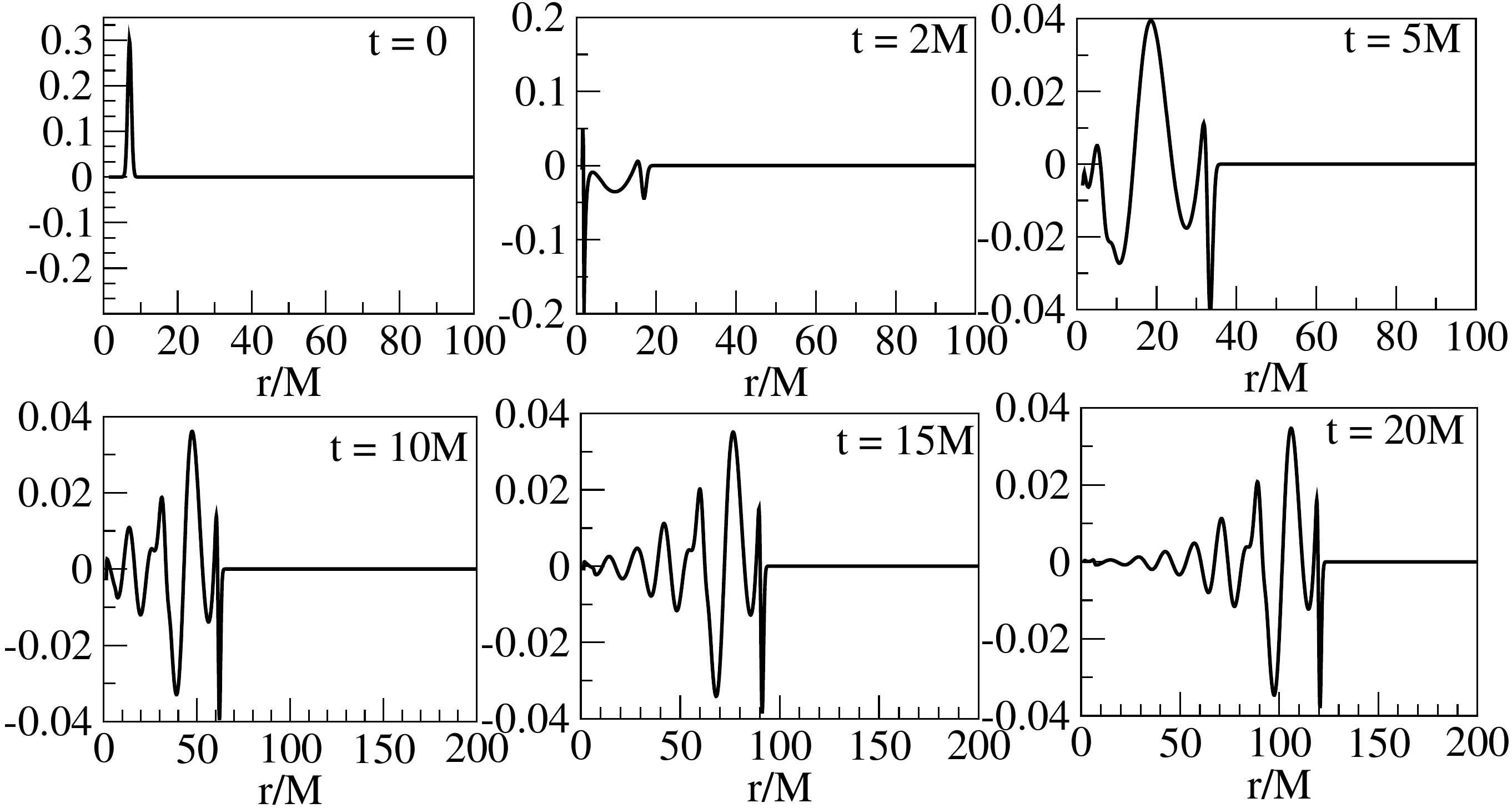} 
\caption{Time evolution of $R^{(1)}_{\pm}$ for $\ell = 2$ for initial data of the form specified in Eq.~(\ref{eq:gauss}). 
Most of the initial pulse falls into the black hole (note the different scales in the vertical axis when passing from 
$t = 0$ to $t = 2M$ to $t = 5M$) after which the spacetime reacts with an oscillating characteristic signal which 
contains the QN modes.}
\label{fig:time_dRN}
\end{figure}

\begin{figure}[h!]
\centering
\includegraphics[height=2.6in]{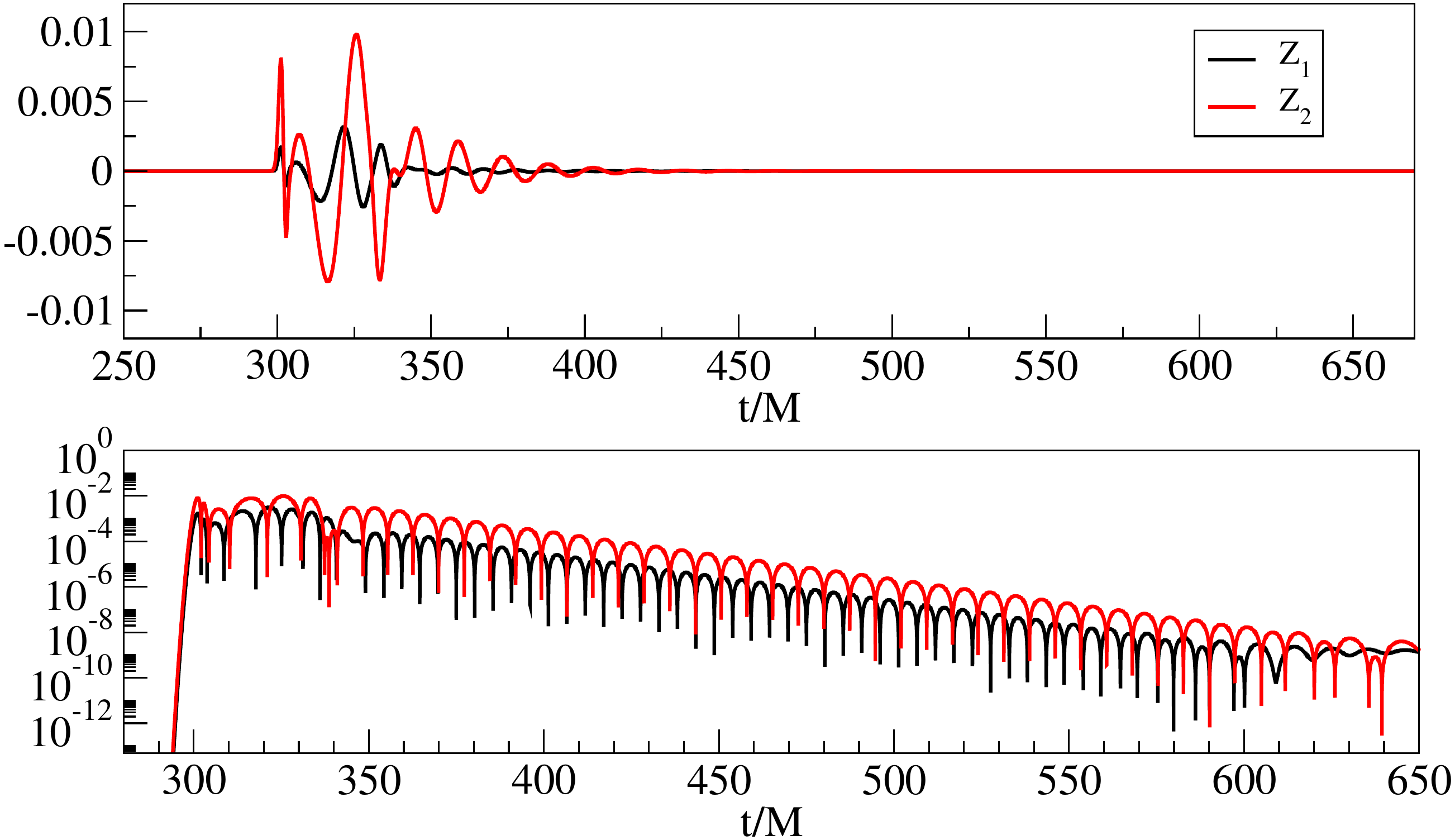} 
\caption{A plot of the signal as a function of time for the ring-down part of the waveform (upper panel). The exponential 
decay rate and the constant frequency are visible in the semi-logarithmic representation (bottom panel).}
\label{fig:log_RN}
\end{figure}

We first solved the evolution equations~(\ref{eq:evolution}) for the case of RN black holes. 
In Fig.~\ref{fig:time_dRN} we display  some snapshots for the time evolution of the amplitude of the perturbation, 
$R^{(1)}_{\pm}$, to illustrate the main properties of the evolution. We have used as initial data a static Gaussian perturbation centered at $r_c$:
\begin{equation}
\label{eq:gauss}
\left. R^{(1)}_{\pm} \right|_{\tilde t = 0} = R_{0} e^{(r-r_{{\rm c}})^2/2\sigma^2},\qquad
\left. \partial_{\tilde t} R^{(1)}_{\pm} \right|_{\tilde t = 0} = 0,
\end{equation}
with $R_{0} = 3\times 10^{-1}$, $r_{c} = 7M$, $\sigma = 0.5M$ and $\left.R^{(2)}_{\pm} \right|_{\tilde t = 0} = 0 =\left.\partial_{\tilde t} 
R^{(2)}_{\pm} \right|_{\tilde t = 0}$.
Solving the system of equations~(\ref{eq:evolution}) one obtains a time series for the perturbation at an observation 
point $r_{obs}$, with $r_{+} < r_{obs} < r_{max}$, where $r_{max}$ is the last point of the numerical grid. In 
Fig.~\ref{fig:log_RN} we plot the functions $Z_1$ and $Z_2$ defined in Eq.~(\ref{eq:transf_z}) 
with $\ell = 2$ for a RN black hole with $Q_m/M = 0.9$. These plots show a clear ringdown signal. Here, 
the observer is located at $ r_{obs} = 300M$, the first point 
in the numerical grid, $r_{+}$, is placed such that it remains inside the event horizon, whereas the last 
point $r_{max}$ is located at $800M$. We have verified that spurious radiation do not affect the signal observed 
at $r = r_{obs}$ for the time we ran our simulation. 

\begin{figure}[h!]
\centering
\includegraphics[height=3.4in]{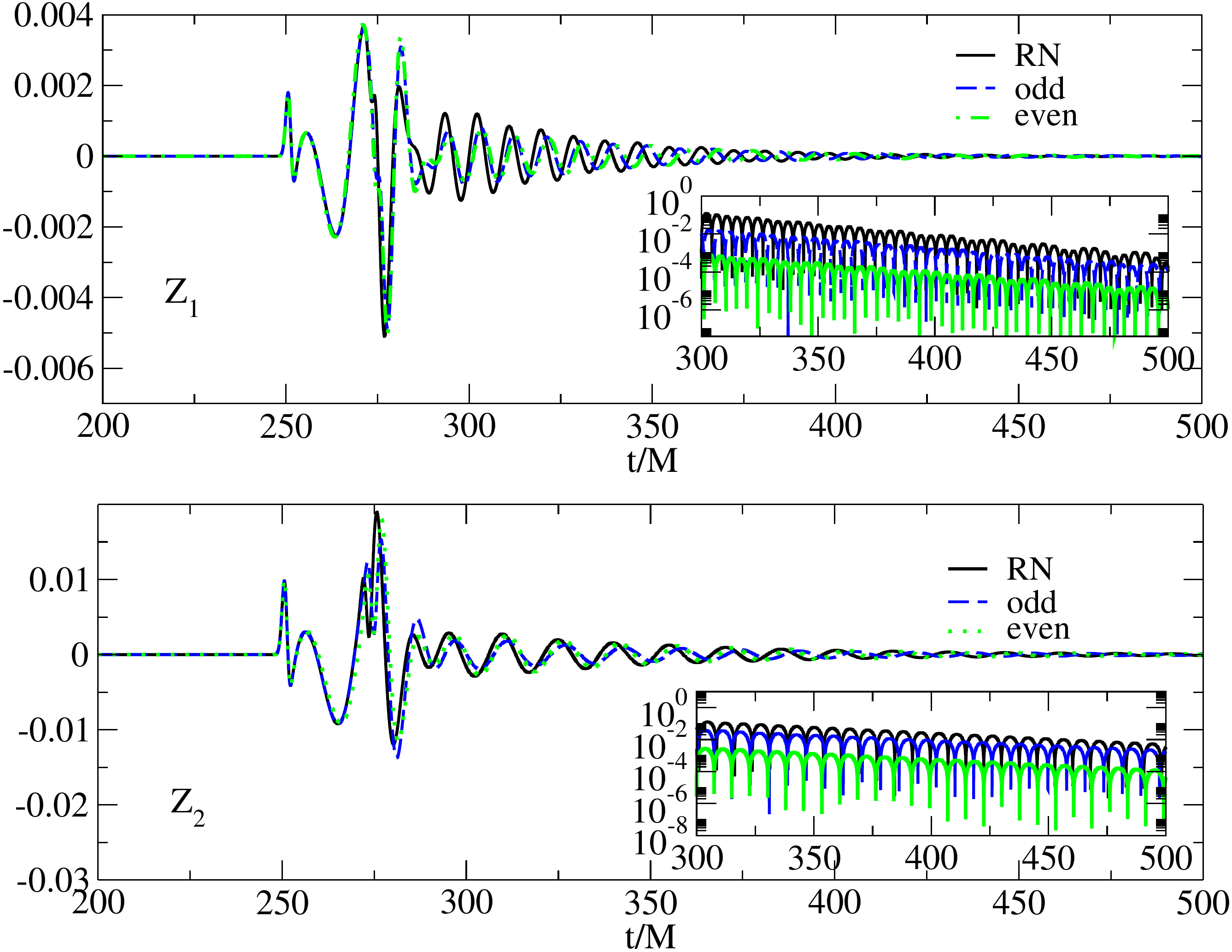} 
\caption{The characteristics of the signals $Z_1(t,r_{obs})$ and $Z_2(t,r_{obs})$ for the deformed RN black holes with $\eta=0.2$ are qualitatively the same as in the undeformed RN case $\eta=0.0$. The insets show the absolute value of the signal in a semi-logarithmic scale to make evident the difference in the frequencies for both parity sectors. For the purpose of the presentation, we have scaled each signal by an appropriate (constant) value.}
\label{fig:time_q08n02}
\end{figure}

Next, we performed time evolutions for different values of the deformation parameter $\eta$, keeping the same parameter 
values for the initial data and for $r_{obs}$. In Fig.~\ref{fig:time_q08n02} we plot the functions $Z_1(t,r_{obs})$ and 
$Z_2(t,r_{obs})$ for $\eta = 0.2$ and $Q_m/M = 0.8$. The small panels show the absolute value of the signals in a 
semi-logarithmic plot in order to better display the ringdown signals and to compare their frequencies to each other. 
From this plot one infers that the frequencies of the quasinormal modes are different than the ones for the RN black 
hole, and further one sees a difference between the two parity sectors. In order to quantify our findings, we performed 
a numerical fit of the functions $Z_1(t,r_{obs})$ and $Z_2(t,r_{obs})$ to a damped exponential, and found the following 
oscillation frequencies: For $Z_1$ we found $M\omega_1 = 0.555$ (even-parity), $M\omega_1 = 0.567$(odd-parity) and for 
$Z_2$, $M\omega_2 = 0.3988$ (even-parity) and $M\omega_2 = 0.402$ (odd-parity). Although in principle our fitting method 
relies on the choice for the window of extraction and the choice of the initial seed to make the fit, we found that these 
frequencies are in good agreement with the ones given in Table~\ref{Tab:freqRNDeformed} below using the more accurate 
method described in the next section. Also, we should mention here that strictly speaking the amplitudes $Z_1$ and $Z_2$ 
defined in Eq.~(\ref{eq:transf_z}) only make sense in the RN case, where the perturbation equations can be decoupled. 
For the deformed RN black holes, it is not possible to decouple the two modes, so that the signal extracted from $Z_1$ 
contains a contribution from the mode oscillating with the frequency $\omega_2$ and the one extracted from $Z_2$ a contribution from the mode with frequency $\omega_1$. 
Nevertheless, one expects this mixing to be small for small values of $\eta$, like to one used in the simulation shown 
in Fig.~\ref{fig:time_q08n02}.

\section{Quasi-normal modes}
\label{Sec-QNM}

In the last section we have shown by means of numerical experiments that an initial perturbation of the black holes described in Sec.~\ref{Sect:EM} gives rise to a clear ringdown signal which contains QN modes. The present section is devoted to a more direct semi-analytic method to compute the QN modes for such black holes which we describe next.

The QN modes are particular solutions of the pulsation equations~(\ref{Eq:PulsationBis}) which are of the form $e^{st}\psi(r)$, with $s = \sigma + i\omega$ a complex number with $\sigma < 0$ and $\psi(r)$ a vector-valued function satisfying the mode equation
\begin{equation}
\left[ s^2 -  N(r)\frac{\partial}{\partial r} N(r)\frac{\partial}{\partial r} 
 + N(r) V_{\ell\pm}(r) \right]\psi(r) = 0,\qquad
r_H < r < \infty.
\label{Eq:ModeEq}
\end{equation}
The QN (complex) frequencies $s$ can be determined as follows (cf. Ref.~\cite{hNbS92} for the case of scalar equations): Let $f_R(s,r)$ be the matrix-valued solution of Eq.~(\ref{Eq:ModeEq}) satisfying the boundary condition
$$
\lim\limits_{r_*\to \infty} e^{s r_*} f_R(s,r) = {\bf I},
$$
for $\re(s) > 0$, with ${\bf I}$ the $2\times 2$ identity matrix and $r_*$ the tortoise coordinate defined in Eq.~(\ref{Eq:Tortoise}). Similarly, let $f_L(s,r)$ be the matrix-valued solution of Eq.~(\ref{Eq:ModeEq}) such that
$$
\lim\limits_{r_*\to -\infty} e^{-s r_*} f_L(s,r) = {\bf I},
$$
for $\re(s) > 0$. The functions $f_{L,R}(s,r)$ are known to be analytic in $s$, so we can consider their analytic continuations to the left complex plane $\re(s) < 0$. The QN frequencies are determined by those values of $s$ for which the Wronskian
\begin{equation}
W(s) := \det\left(\begin{array}{rr}
 f_L(s,r) & f_R(s,r) \\
 \frac{\partial f_L}{\partial r_*}(s,r) & \frac{\partial f_R}{\partial r_*}(s,r)
\end{array} \right),
\end{equation}
vanishes. When $W(s) = 0$, there are constant vectors $(a,b)\in \Complex^2$ and $(c,d)\in \Complex^2$ such that
$$
\psi(s,r) := f_L(s,r)\left( \begin{array}{c} a \\ b \end{array} \right)
 =  f_R(s,r)\left( \begin{array}{c} c \\ d \end{array} \right),
$$
and $e^{st}\psi(s,r)$ is the corresponding QN mode. Because $\re(s) < 0$, such a mode is damped in time, although one should note that it diverges as $r_*\to \pm\infty$ along surfaces of constant time $t$.

\subsection{Numerical computation method}
\label{SubSec:NumericalMethod}

The analytic continuations of the two solutions $f_{L,R}(s,r)$ of Eq.~(\ref{Eq:ModeEq}) for $\im(s) > 0$ can be constructed using a straightforward generalization of the method recently described in Ref.~\cite{eCmMoS15} for scalar equations. This methods consists in approximating the function $f_R(s,r)$ via an iteration scheme,
\begin{equation}
f_R(s,r) = e^{-s r_*}\lim\limits_{k\to \infty} (T_{R,s}^k {\bf I})(r),
\label{Eq:fRIter}
\end{equation}
with the operator $T_{R,s}$, acting on continuous and bounded matrix-value functions $\xi$, defined by
\begin{equation}
(T_{R,s}\xi)(r) := {\bf I} + \frac{1}{2s}\int\limits_{\gamma_\alpha} \left[
  1 - \exp\left( -2s\int\limits_r^{r'} \frac{dr''}{N(r'')} \right) \right] V_{\pm\ell}(r')\xi(r') dr',\qquad
\re(r) > r_H,
\label{Eq:TRs}
\end{equation}
where it is understood that the integral from $r$ to $r'$ in the exponential is performed along the path $\gamma_\alpha$. Here, the integration path $\gamma_\alpha$ is the following ray in the complex $r$-plane:
$$
\gamma_\alpha(\lambda) = r + e^{i\alpha}\lambda,\qquad \lambda\geq 0,
$$
with $\alpha$ an angle chosen to be slightly larger than $-\pi/2$. As argued in~\cite{eCmMoS15} the integral converges for all $s = |s| e^{i\varphi}$ with $-\pi/2 < \alpha + \varphi < \pi/2$, so choosing $\alpha = -\pi/2$ yields convergence for all $\im(s) > 0$.

Likewise, the solution $f_L(s,r)$ can be obtained from
\begin{equation}
f_L(s,r) = e^{+s r_*}\lim\limits_{k\to \infty} (T_{L,s}^k {\bf I})(r),
\label{Eq:fLIter}
\end{equation}
with the operator $T_{L,s}$, acting on continuous and bounded matrix-value function $\xi$, defined by
\begin{equation}
(T_{L,s}\xi)(r) := {\bf I} - \frac{1}{2s}\int\limits_{\Gamma_\beta} \left[
  1 - \exp\left( 2s\int\limits_r^{r'} \frac{dr''}{N(r'')} \right) \right] V_{\pm\ell}(r')\xi(r') dr',
\label{Eq:TLs}
\end{equation}
with the curve
$$
\Gamma_\beta(\lambda) = r_H + (r - r_H)\exp( -e^{i\beta}\lambda ),\quad \lambda \geq 0,
$$
with $\beta$ slightly larger than $-\pi/2$, which spirals counter-clockwise around the point $r = r_H$ in the complex $r$-plane.

For a given value of $s$ such that $\im(s) > 0$ we numerically compute the functions $f_{L,R}(s,r)$ and their first derivatives by truncating the iterations in Eqs.~(\ref{Eq:fRIter},\ref{Eq:fLIter}) at some finite $k$ and approximating the integral operators in Eqs.~(\ref{Eq:TRs},\ref{Eq:TLs}) using the trapezoidal rule with $n$ grid-points. We choose $\alpha = -1.57$ and $\beta = -1.5$ and find that in practice, of the order of $n\sim 5\times 10^4$ points but only about $k\sim 20$ iterations are needed in order to achieve a reasonable accuracy. Finally, the zeros of the Wronski determinant are determined numerically using a standard Newton algorithm~\cite{PFTV86}.

\subsection{Eikonal limit}
\label{SubSec:Eikonal}

In the high-frequency approximation, the 
QN frequencies can be understood directly from the properties of the unstable circular 
null geodesics of the underlying spacetime, see~\cite{vCaMeBhWvZ09,eBvCaS09} 
and references therein. Indeed, in this limit the QN oscillations can be interpreted in 
terms of decaying wave packets which are localized along an unstable circular null geodesic. 
Denoting by $\Omega_{circ}$ and $\Lambda_{circ}$ the angular velocity and the Lyapunov exponent, respectively, of the unstable null geodesic, the QN frequencies for high values of $\ell$ are given by the formula~\cite{vCaMeBhWvZ09} 
\begin{equation}
Ms = -\left( n_o + \frac{1}{2} \right)M\Lambda_{circ} + i\ell\,M\Omega_{circ},
\label{Eq:QNMEikonalSpectrum}
\end{equation}
with $n_o$ the overtone number. An unstable circular null geodesics is characterized by local maxima of the function
\begin{equation}
H(r) := \frac{1}{r^2}\left( 1 - \frac{2m(r)}{r} \right).
\label{Eq:HDef}
\end{equation}
If $r = r_{circ}$ is the location of such a maximum, then, the associated angular velocity and Lyapunov exponent are given by~\cite{vCaMeBhWvZ09}
\begin{equation}
\Omega_{circ} = \left.  \sqrt{H(r)} \right|_{r = r_{circ}},\quad
\Lambda_{circ} = \left. \frac{1}{\sqrt{2}}\left( 1 - \frac{2m(r)}{r} \right) 
 \sqrt{ -\frac{1}{H(r)} \frac{d^2}{dr^2} H(r) } \right|_{r = r_{circ}}.
\label{Eq:OmegaLambda}
\end{equation}

Before computing the eikonal limit for the deformed black holes, it is important to realize that the geometric optics approximation of NED has a much richer structure than in the Maxwell case. In particular, it is not necessarily true anymore that light rays propagate along null geodesics of the spacetime metric ${\bf g}$. Instead, the light rays are described by co-vector fields ${\bf k} = d\Psi$ with $\Psi$ the phase function, which obeying a quartic Fresnel equation, see for instance~\cite{yOgR02,fAfCeGoR15}. For the particular models of NED considered in the present article, this quartic equation factorizes into a set of two conditions
\begin{equation}
\mathfrak{g}_1^{\mu\nu} k_\mu k_\nu = 0,\qquad
\mathfrak{g}_2^{\mu\nu} k_\mu k_\nu = 0,
\end{equation}
where the reciprocal effective metrics are given by~\cite{yOgR02,fAfCeGoR15}
\begin{equation}
\mathfrak{g}_1^{\mu\nu} = g^{\mu\nu} + b_1 F^{\mu}{}_{\alpha} F^{\nu\alpha},\qquad
\mathfrak{g}_2^{\mu\nu} = g^{\mu\nu},
\label{Eq:EffectiveMetrics}
\end{equation}
with
$$
b_1 = \frac{{\cal L}_{FF}}{{\cal L}_F} = (\kappa-1)\frac{g^2}{y^2}.
$$
Therefore, and interestingly, the light propagation is characterized by a set of two cones at each event, one coinciding with the cone described by the spacetime metric, and the other one described by the effective metric $\mathfrak{g}_1$ which depends on the nonlinear electromagnetic field. The presence of this second cone leads to interesting effects including birefringence~\cite{yOgR02}. Also, the fact of having two cones plays an important role in the Cauchy problem for the nonlinear electromagnetic field: it has recently been shown in~\cite{fAfCeGoR15} that a given NED theory admits a symmetric hyperbolic formulation if and only if the interior of the two cones have a nonempty intersection.

For the particular solutions considered in this article, one finds
$$
F^{\mu}{}_{\alpha} F^{\nu\alpha}
\frac{\partial}{\partial x^\mu}\otimes \frac{\partial}{\partial x^\nu}
 = \frac{Q_m^2}{r^6}\left( \frac{\partial}{\partial \vartheta}\otimes \frac{\partial}{\partial \vartheta}
 + \frac{1}{\sin^2\vartheta}\frac{\partial}{\partial \varphi}\otimes \frac{\partial}{\partial \varphi}
 \right),
$$
and thus the reciprocal effective metric $\mathfrak{g}_1^{-1}$ is
\begin{equation}
\mathfrak{g}_1^{-1}
 =  -\left( 1 - \frac{2m(r)}{r} \right)^{-1}\frac{\partial}{\partial t}\otimes \frac{\partial}{\partial t}
 + \left( 1 - \frac{2m(r)}{r} \right)\frac{\partial}{\partial r}\otimes \frac{\partial}{\partial r}
 + \frac{\kappa(r)}{r^2} \left( \frac{\partial}{\partial \vartheta}\otimes \frac{\partial}{\partial \vartheta}
 + \frac{1}{\sin^2\vartheta}\frac{\partial}{\partial \varphi}\otimes \frac{\partial}{\partial \varphi}
 \right).
\label{Eq:EffMetric1}
\end{equation}
The relevant properties of the unstable, circular null geodesics associated with this effective metric are described by the same formulae as Eq.~(\ref{Eq:OmegaLambda}), except that the function $H(r)$ needs to be replaced with
$$
H_1(r) = \frac{\kappa(r)}{r^2}\left( 1 - \frac{2m(r)}{r} \right).
$$
The radius of the circular null geodesics as well as the associated quantities $\Omega_{circ}$ and $\Lambda_{circ}$ will be computed below for the RN metric (in which case $\mathfrak{g}_1 =\mathfrak{g}_2 = {\bf g}$) and in the next section for deformed RN black holes in NED, for which $\mathfrak{g}_1 \neq \mathfrak{g}_2$.

\subsection{Tests for RN black holes}

In order to test our code, we first apply it to the RN case for which
$$
N(r) = 1 - \frac{2M}{r} + \frac{Q_m^2}{r^2}.
$$
The corresponding expressions for the effective potentials $V_{\ell\pm}$ are given in Eqs.~(\ref{Eq:EffPotRN-},\ref{Eq:EffPotRN+}), and as discussed in Sec.~\ref{SubSec:DecouplingRN} in each parity sector, the pulsation equations can be decoupled into two scalar wave equations for the gauge-invariant quantities $Z^{(\pm)}_i$, $i=1,2$, which simplifies the problem considerably. However, to test the validity of our method described in the previous section and also in view of the applications presented in the next section regarding the deformed RN black holes, here we do not make use of this decoupling.

For the numerical implementation of our method we find it convenient to work with the dimensionless quantities
$$
x := \frac{r}{r_H},\qquad 
\overline{M} := \frac{M}{r_H},\qquad
\overline{Q}_m := \frac{Q_m}{r_H},\qquad
q := \frac{Q_m}{M},
$$
such that the event horizon is located at $x = 1$. Since $0 = N(r_H) = 1 - 2\overline{M} + q^2\overline{M}^2$ we find the following expressions for the dimensionless total mass and magnetic charge:
$$
\overline{M} = \frac{1}{1 + \sqrt{1 - q^2}},\qquad
\overline{Q}_m = \frac{q}{1 + \sqrt{1 - q^2}}.
$$

\begin{figure}[h!]
\centering
\includegraphics[height=3.0in]{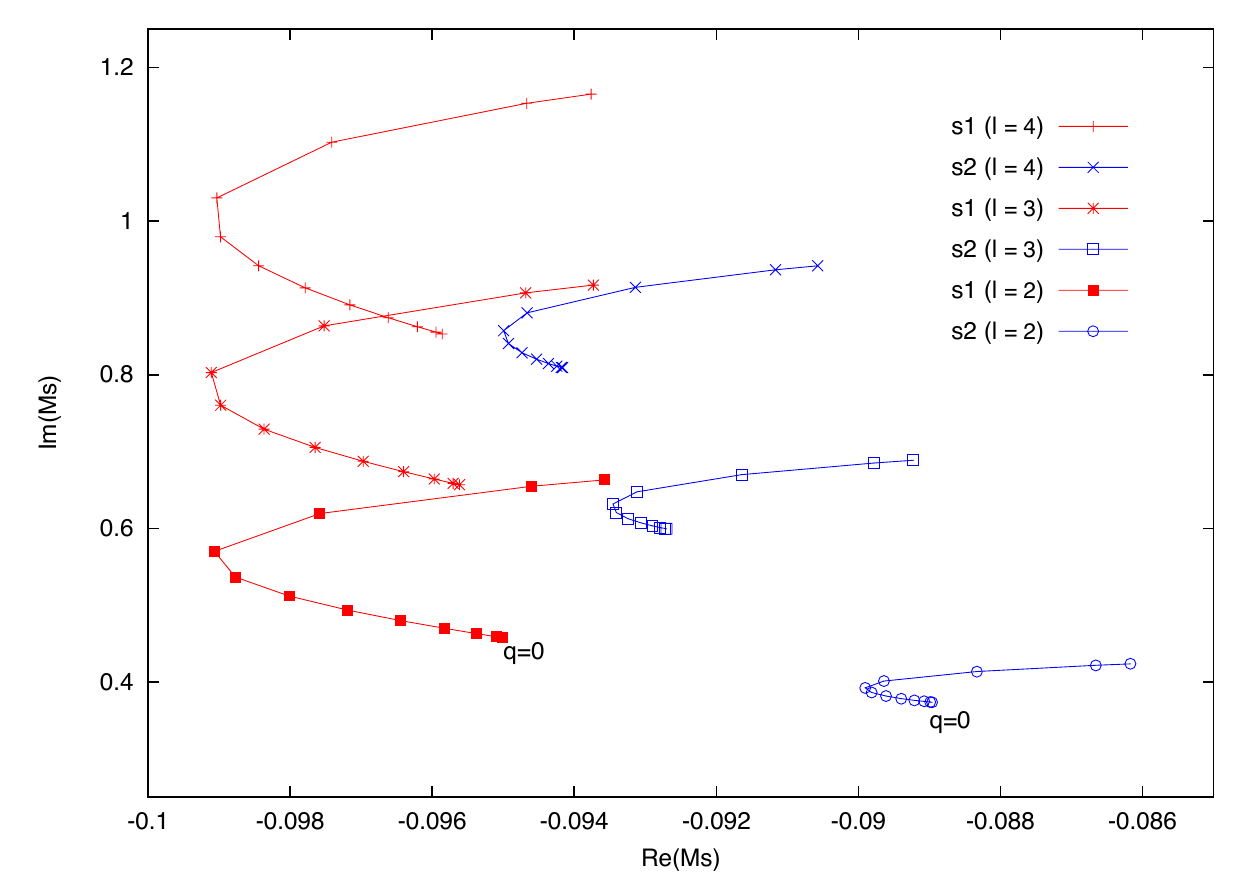}  
\caption{RN fundamental QN frequencies for different values of $q$ and angular momenta $\ell = 2,3,4$. For $\ell=2$ the point corresponding to $q=0$ is indicated in the figure. Subsequent points correspond to the values $q = 0.1,0.2,0.3,0.4,0.5,0.6,0.7,0.8,0.9,0.95,0.96$.}
\label{Fig:RNSpirals}
\end{figure}

Using our numerical method described in Sec.~\ref{SubSec:NumericalMethod} we computed the fundamental QN frequencies $s$ for angular momenta $\ell = 2,3,\ldots,9$ for RN black holes with different charge to mass ratio $q$. Results are shown in Tables~\ref{Tab:freqRN} and~\ref{Tab:QNM_RN} and in Fig.~\ref{Fig:RNSpirals}. As mentioned previously, each parity sector gives rise to two families of QN modes corresponding to the two quantities $Z_1^{(\pm)}$ and $Z_2^{(\pm)}$ defined in Sec.~\ref{SubSec:NumericalMethod}. We denote the corresponding frequencies by $s_1^{(\pm)}$ and $s_2^{(\pm)}$, respectively. Recall that in the RN case there exists a symmetry between the two parity sectors which implies that $s_1^{(+)} = s_1^{(-)} =: s_1$ and $s_2^{(+)} = s_2^{(-)} =: s_2$, so that the superscript $^{(\pm)}$ is superfluous in this case. In order to check this symmetry, we have computed the frequencies in both parity sectors independently with our numerical method, and found that the results agree (within the precision of our code) with each other. Furthermore, we have checked that our frequencies are consistent with those given in standard references, see for instance Table~V in chapter 5 of Ref.~\cite{Chandrasekhar-Book}.

\begin{table}[h]
\begin{tabular}{|l|c|c|c|c|} 
 \hline
 $q$ & $M s_1$ ($\ell = 2$)  & $M s_2$ ($\ell = 2$) &  $M s_1$ ($\ell = 3$) & $M s_2$ ($\ell = 3$)\\ 
\hline
$\,0.0$  & $\,-0.0950045 + 0.457596i\,$  &   $\,-0.0889623 + 0.373672i\,$ & $\,-0.0956164 + 0.656899i\,$  &  $\,-0.0927032 + 0.599444i\,$ \\   
$0.2$ & $-0.0953735 + 0.462965i$ &   $-0.0890748 + 0.374745i$     & $-0.0959735 + 0.664367i$  &  $-0.0927897 + 0.601029i$\\
$0.4$  & $-0.0964422 + 0.479926i$ &   $-0.0893981 + 0.378437i$  & $-0.0969736 + 0.687281i$ &  $-0.0930587 + 0.607057i$  \\
$0.6$  & $-0.0980168 + 0.512011i$ &   $-0.0898137 + 0.386218i$ & $-0.0983684 + 0.729188i$   &  $-0.0934102 + 0.620661i$  \\  
$0.8$ &   $-0.0896432 + 0.401217i$   & $-0.0990692 + 0.570131i$  &  $-0.0931173 + 0.647552i$   & $-0.0991131 + 0.802845i$  \\
$0.9$  & $-0.0975831 + 0.619398i$   &   $-0.0883330 + 0.413571i$ & $-0.0975242 + 0.863759i$ &  $-0.0916440 + 0.670024i$  \\  
$0.95$   & $-0.0946052 + 0.654763i$ &  $-0.0866585 + 0.421693i$  & $-0.0946854 + 0.906681i$ &  $-0.0897801 + 0.685188i$ \\
 \hline
\end{tabular}
\caption{Fundamental QN frequencies for RN black holes with charge $Q_m = q M$ and angular momentum number $\ell=2,3$. For each values of $q$ and $\ell$ there is a pair of fundamental frequencies $(s_1,s_2)$, as explained in the text.}
\label{Tab:freqRN}
\end{table}

In the eikonal limit discussed in Sec.~\ref{SubSec:Eikonal} we find the following properties for the unstable circular null geodesic in the RN case (cf.~\cite{sFjC12}): its radius is
$$
r_{circ} = \frac{3M}{2}\left(1+\sqrt{1- \tfrac{8}{9}q^2}\right),\qquad
q = \frac{Q_m}{M},
$$
and the associated angular frequency and Lyapunov exponent are given by
\begin{equation}
(M\Omega_{circ})^2 = \frac{2}{27}\frac{1 + 3\sqrt{1-\tfrac{8}{9}q^2}}
{\left(1+\sqrt{1- \tfrac{8}{9}q^2}\right)^3},\qquad
(M\Lambda_{circ})^2 = \frac{4}{27}
\frac{\left(1+3\sqrt{1-\tfrac{8}{9}q^2} \right)\sqrt{1-\tfrac{8}{9}q^2}}
{\left(1+\sqrt{1- \tfrac{8}{9}q^2}\right)^4}.
 \end{equation}
 For example, when $q = 0.8$ we obtain the values $M\Omega_{circ} \simeq 0.219974$ and $ -\Lambda_{circ} M/2 \simeq -0.0979256$. We found that these values agree fairly well with the numerical computed frequencies for $q = 0.8$ and high values of $\ell$, see Table~\ref{Tab:QNM_RN}.

\begin{table}[h]
\center
\begin{tabular}{| r | c | c | c | c |}
\hline
$\ell$ & $M s_1$  & $M s_2$ & $\Delta(\im(M s_1))$ & $\Delta(\im(M s_2))$ \\  
\hline
$2$ &  $ -0.0990692 + 0.570131i$ & $ -0.0896433 + 0.401217i$  & & \\
$3$ & $   -0.0991127 + 0.802844i$ & $  -0.0931173 + 0.647552i$  & $0.2327$ & $0.2463$  \\
$4$ & $-0.0990335 + 1.030399i$ &  $-0.0946653 + 0.880566i$  & $0.2276 $ & $0.2330$  \\
$5$  &  $-0.0989373 + 1.255416i$  & $-0.0954892 + 1.108314i$ & $0.2250$ &  $0.2277$   \\
$6$ & $-0.0988473 + 1.478992i$ & $ -0.0959911 + 1.333450i$ & $0.2236$ & $0.2251$  \\ 
$7$ & $-0.0987682 + 1.701668i$ &   $-0.0963263  +1.557103i$  & $0.2227$ & $0.2237$   \\
$8$ & $-0.098698  + 1.92374i$ &  $-0.096564 +   1.77983 i$ & $0.2221$ & $0.2227$  \\
$9$ & $-0.09864  +  2.14540i$ & $-0.09674  +  2.00194 i$  &  $0.2217$ & $0.2221$ \\
\hline
\end{tabular}
\caption{Fundamental QN frequencies for a RN black hole with angular momentum number $\ell=2,3,...,9$ and charge to mass ratio $q=0.8$. The fourth  column shows the difference of $\im(M s_1)$ between two successive values of $\ell$'s, and likewise for the fifth column. Note that these values are consistent with the limiting value $M\Omega_{circ} \simeq 0.219974$ predicted by the eikonal limit, while the real parts of the frequencies are consistent with the value $-\Lambda_{circ} M/2\simeq -0.0979256$ predicted for $\ell\to \infty$.}
\label{Tab:QNM_RN}
\end{table}

\section{Results for the deformed RN black holes}
\label{Sec:Results}

In this section we apply our numerical method described and tested in the previous section to the computation of QN frequencies for the deformed RN black holes discussed in Sec.~\ref{SubSec:Alternative} and Appendix~\ref{App:SolutionProp}. We recall that relevant explicit expressions for the metric and the effective potentials are summarized in Sec.~\ref{SubSec:Explicit}, and that in the limit $\eta \to 0$ these expressions reduce to the RN solution.

As for the RN tests in the previous section, we find it convenient to work with the dimensionless quantities
$$
x := \frac{r}{r_H},\qquad 
\overline{M} := \frac{M}{r_H},\qquad
\overline{Q}_m := \frac{Q_m}{r_H},\qquad
\overline{\eta} := \frac{\eta}{r_H},\qquad
q := \frac{Q_m}{M},
$$
such that the event horizon is located at $x = 1$. Since
$$
0 = N(r_H) = 1 - 2\overline{M} + q^2\overline{M}^2 E(\overline{\eta}),\qquad
E(\overline{\eta}) := \frac{1}{2}\left[ \frac{1}{1 + \overline{\eta}^2} + \frac{\arctan(\overline{\eta})}{\overline{\eta}} \right],
$$
one finds the expressions
$$
\overline{M} = \frac{1}{1 + \sqrt{1 - E(\overline{\eta}) q^2}},\qquad
\overline{Q}_m = \frac{q}{1 + \sqrt{1 - E(\overline{\eta}) q^2}},
$$
in terms of of the charge to mass ratio $q$ and the dimensionless deformation parameter $\overline{\eta}$.

\subsection{Fundamental frequencies for the deformed RN black holes}

\begin{figure}[h!]
\centering
\includegraphics[height=3.0 in]{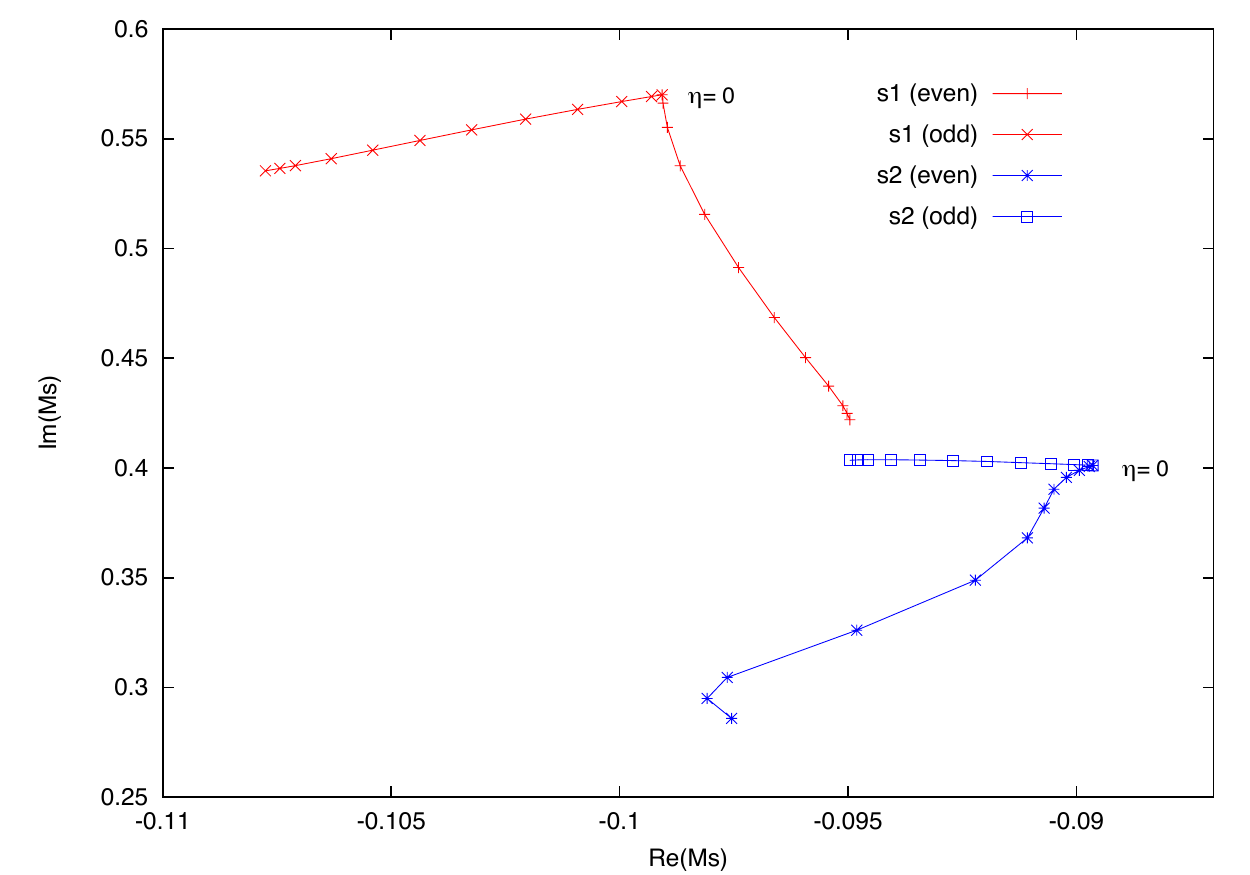}  
\caption{Fundamental quadrupolar ($\ell=2$) QN frequencies for deformed RN black holes with $q = 0.8$ and varying $\eta$. The two frequencies corresponding to the undeformed RN case are indicated by $\eta=0$ in the figure. Subsequent points correspond to the values $\eta = 0.1,0.2,0.3,0.4,0.5,0.6,0.7,0.8,0.9,0.95,1.0$.}
\label{Fig:DeformedRN1}
\end{figure}

\begin{figure}[h!]
\centering
\includegraphics[height=3.0in]{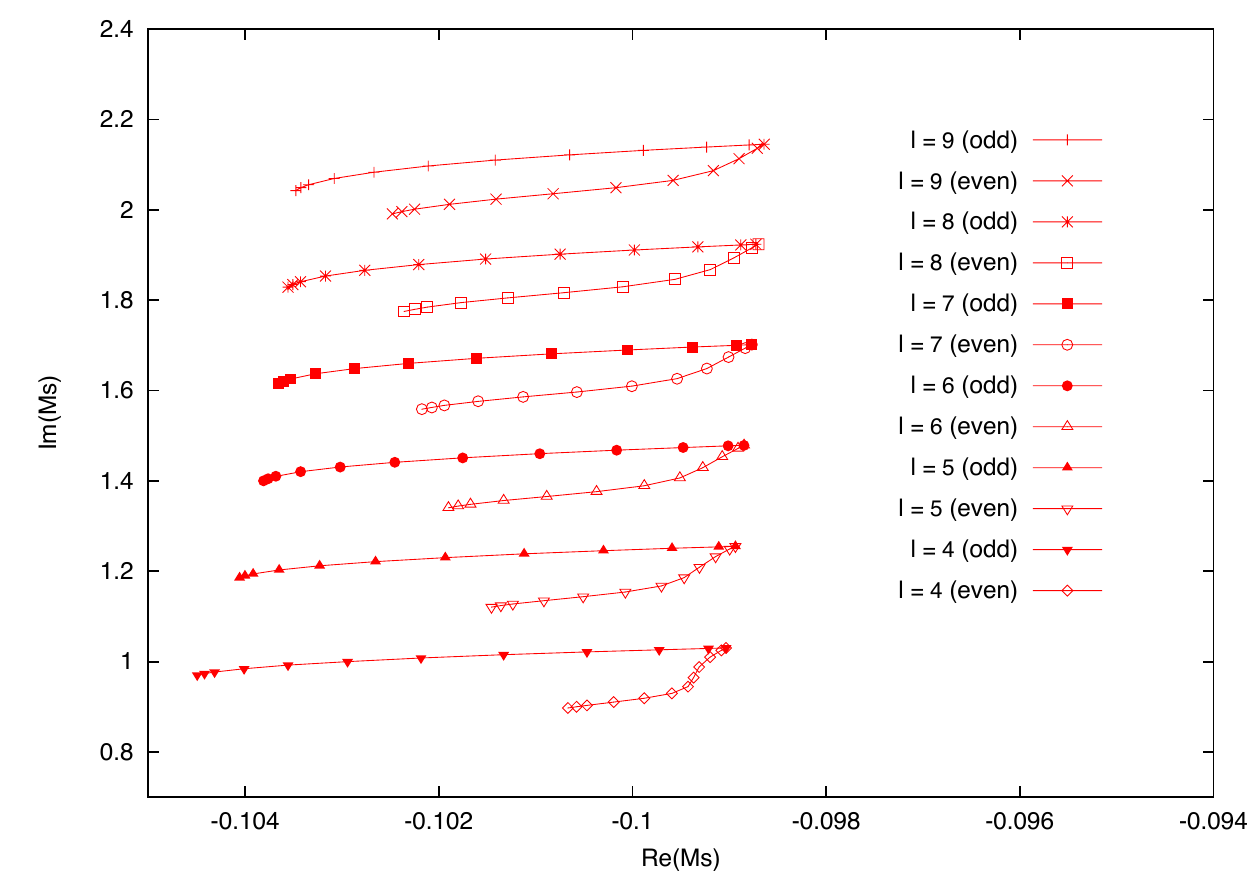}
\caption{Fundamental QN frequencies $s_1$ for the deformed RN black holes with $q = 0.8$ in both parity cases and $\ell=4,5,\ldots,9$. For each value of the angular momentum number $\ell$, we vary $\eta$ between $0.0$ and $1.0$, as in the previous figure.}
\label{Fig:Splitting1}
\end{figure}

\begin{figure}[h!]
\centering
\includegraphics[height=3.0in]{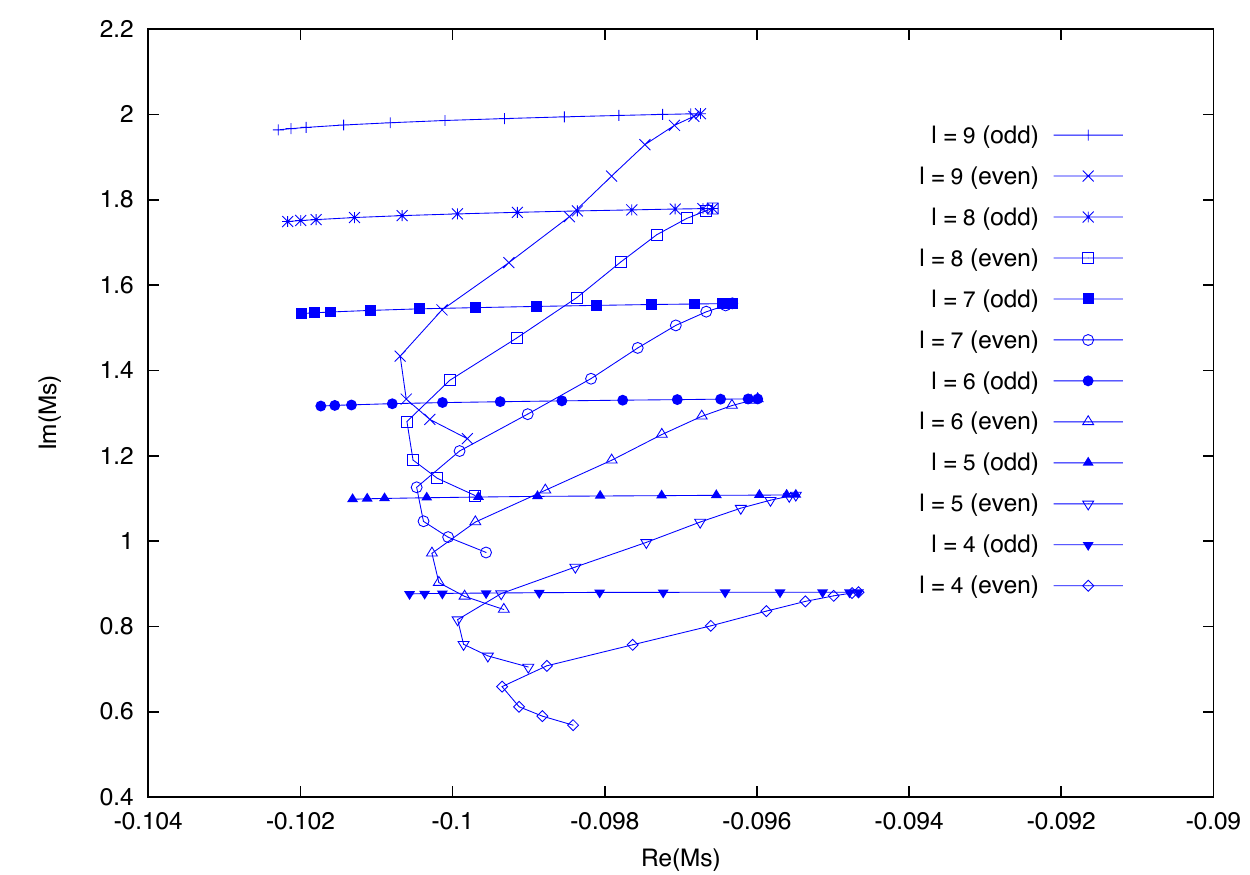}  
\caption{Fundamental QN frequencies $s_2$ for the deformed RN black holes with $q = 0.8$ in both parity cases and $\ell=4,5,\ldots,9$. For each value of the angular momentum number $\ell$, we vary $\eta$ between $0.0$ and $1.0$, as in the previous figure.}
\label{Fig:Splitting2}
\end{figure}

For the deformed RN black holes, we compute numerically the QN fundamental frequency pair $(s_1^{(\pm)},s_2^{(\pm)})$ in each parity sector. Contrary to the RN case, where the results in both parity sectors agree with each other, we found that in the deformed case there is a splitting between the two parity sectors which becomes more important the larger the deformation is. This splitting phenomena is shown in Table~\ref{Tab:freqRNDeformed} and illustrated in Figs.~\ref{Fig:DeformedRN1},\ref{Fig:Splitting1} and~\ref{Fig:Splitting2}.

\begin{table}[h]
\begin{tabular}{|r|c|c|c|c|} 
 \hline
 $\eta\,\,$   & $M s_1$ (even) & $M s_1$ (odd) &  $M s_2$ (even) & $M s_2$ (odd)  \\ 
\hline \hline
$\,0.0$   & $\,-0.09906914 + 0.570130i\,$  & $\,-0.09906914 + 0.570130i\,$  &  
$\,-0.08964324 + 0.401217i\,$ & $\,-0.08964322 + 0.401217i\,$  \\   
$0.10$   & $-0.09905244 + 0.566328i$   & $-0.09930090 + 0.569329i$  &  
$-0.08972280 + 0.400673i$  & $-0.08974988 + 0.401315i$  \\
$0.20$   & $-0.0989559 + 0.555218i$     & $-0.09995440 + 0.567014i$  &  
$-0.08993771 + 0.398944i$ & $-0.09006426 + 0.401593i$  \\
$0.30$   & $-0.09867505 + 0.537735i$   & $-0.100920 + 0.563440i$   &  
$-0.09022224 + 0.395718i$ & $-0.09056605 + 0.402005i$  \\  
$0.40$   & $-0.09814032 + 0.515589i$   & $-0.102058 + 0.558997i$   &  
$-0.09049119 + 0.390355i$ & $-0.09121493 + 0.402487i$  \\
$0.50$   & $-0.09739969 + 0.491403i$   & $-0.103239 + 0.554137i$   & 
$-0.09071092 + 0.381708i$ & $-0.09195216 + 0.402968i$  \\  
$0.60$   & $-0.09661410 + 0.468616i$   & $-0.104372 + 0.549295i$   &  
$-0.09107389 + 0.368161i$ & $-0.09271078 + 0.403382i$  \\
$0.70$   & $-0.09593174 + 0.450317i$   & $-0.105404 + 0.544824i$   &  
$-0.09221340 + 0.348875i$ & $-0.09342863 + 0.403676i$  \\
$0.80$   & $-0.09542468 + 0.437319i$   & $-0.106314 + 0.540961i$   &  
$-0.09480851 + 0.326114i$ & $-0.09405807 + 0.403814i$ \\
$0.90$   & $-0.09511355 + 0.428376i$   & $-0.107096  + 0.537836i$  &  
$-0.09764332 + 0.304574i$ & $-0.09457018 + 0.403778i$ \\
$0.95$   & $-0.09502128 + 0.424942i$   & $-0.107440  + 0.536562i$  &  
$-0.09808502 + 0.294981i$ & $-0.09477820 + 0.403694i$ \\
$1.00$   & $-0.09496011 + 0.422005i$   & $-0.107754 + 0.535480i$   &  
$-0.09755027 + 0.285906i$ & $-0.09495402 + 0.403566i$ \\
 \hline
\end{tabular}
\caption{Fundamental QN frequencies with $\ell=2$ for the deformed RN black holes with charge $q = 0.8$ and $\eta = 0.1,0.2,0.3,0.4,0.5,0.6,0.7,0.8,0.9,0.95,1.0$.}
\label{Tab:freqRNDeformed}
\end{table}

\subsection{Eikonal limit}

In this subsection, we provide a partial explanation for the parity splitting phenomena we encountered when 
considering the deformed RN black hole in NED. To this purpose, we recall the discussion in Sec.~\ref{SubSec:Eikonal} 
where we pointed out that in the geometric optics approximation, electromagnetic waves in NED propagate on 
null geodesics with respect to either one of the two effective metrics $\mathfrak{g}_1$ and $\mathfrak{g}_2$ given 
in Eq.~(\ref{Eq:EffectiveMetrics}). For our model, $\mathfrak{g}_2$ agrees with the spacetime metric while 
$\mathfrak{g}_1$ is explicitly given by
Eq.~(\ref{Eq:EffMetric1}). For our alternative model, we have
$$
H_1(r) = \frac{\kappa(r)}{r^2}\left[
 1 - \frac{2M}{r} + \frac{Q_m^2}{2r}\left( \frac{r}{r^2 + \eta^2} + \frac{\arctan(\eta/r)}{\eta} \right)
 \right],\qquad \kappa(r) = \frac{1 - \frac{2\eta^2}{r^2}}{1 + \frac{\eta^2}{r^2}},
$$
and for the effective metric $\mathfrak{g}_2$ the function $H_1(r)$ is given by the same expression with $\kappa(r)$ set to one.

\begin{figure}[h!]
\centering
\includegraphics[height=2.4in]{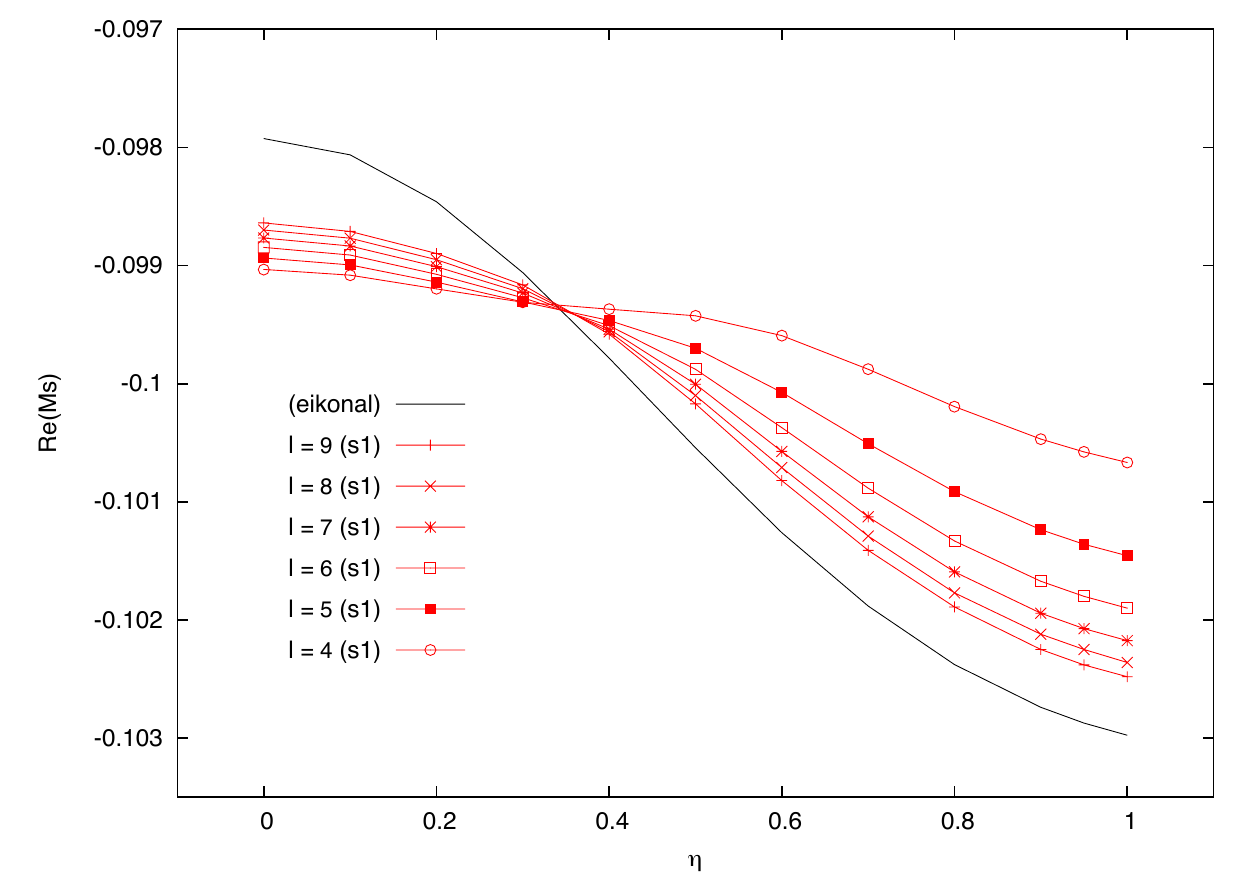} 
\includegraphics[height=2.4in]{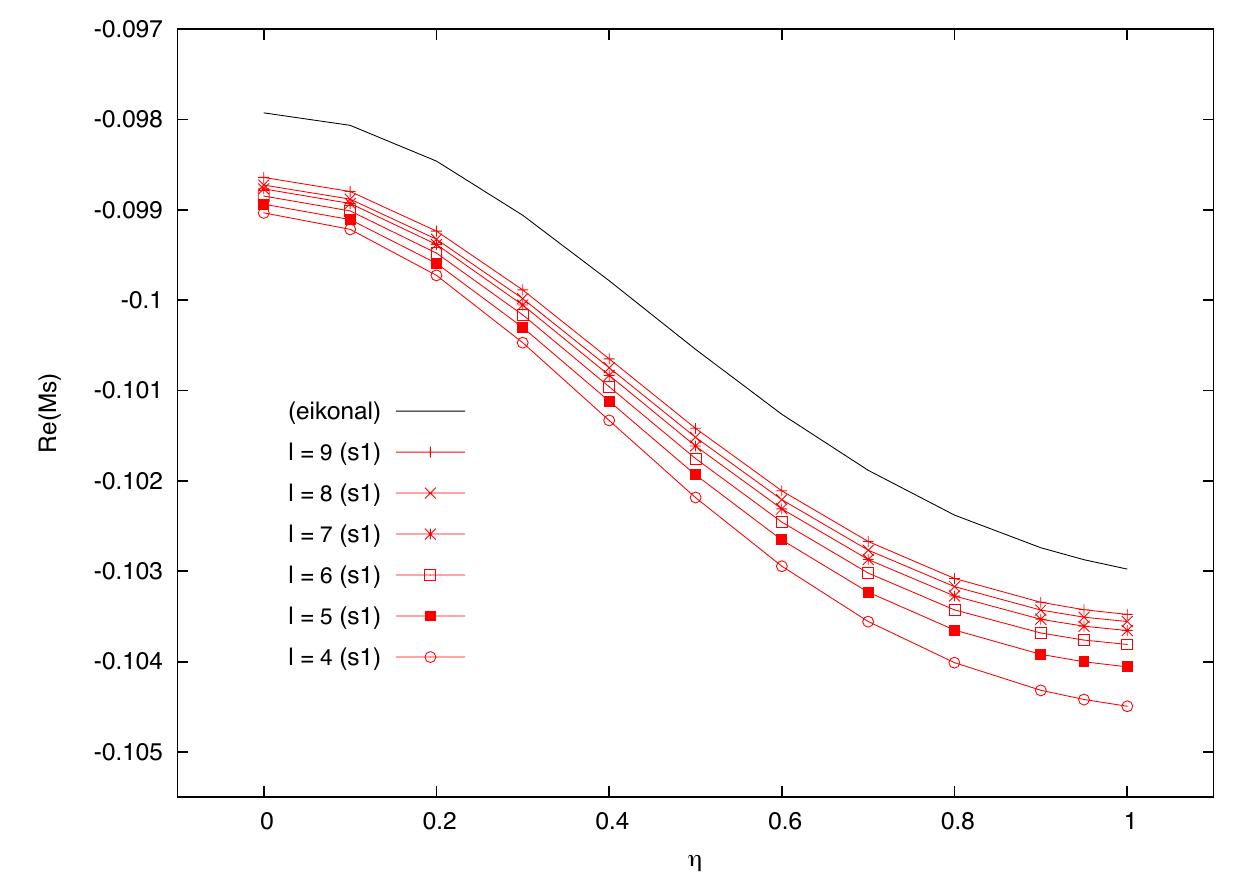}  
\caption{Decay rates for the frequency $s_1$ as a function of the deformation parameter $\eta$ and the corresponding result from the eikonal limit indicated here by the solid black line.
Left panel: Even parity sector, and the eikonal limit from the effective metric $\mathfrak{g}_2$. Right panel: Odd parity sector, and the eikonal limit from the effective metric $\mathfrak{g}_2$. As we see, for large values of $\ell$ the behavior in each parity sector is consistent with the predictions from the eikonal limit for the metric $\mathfrak{g}_2$.}
\label{Fig:Def1ReEta}
\end{figure}

\begin{figure}[h!]
\centering
\includegraphics[height=2.4in]{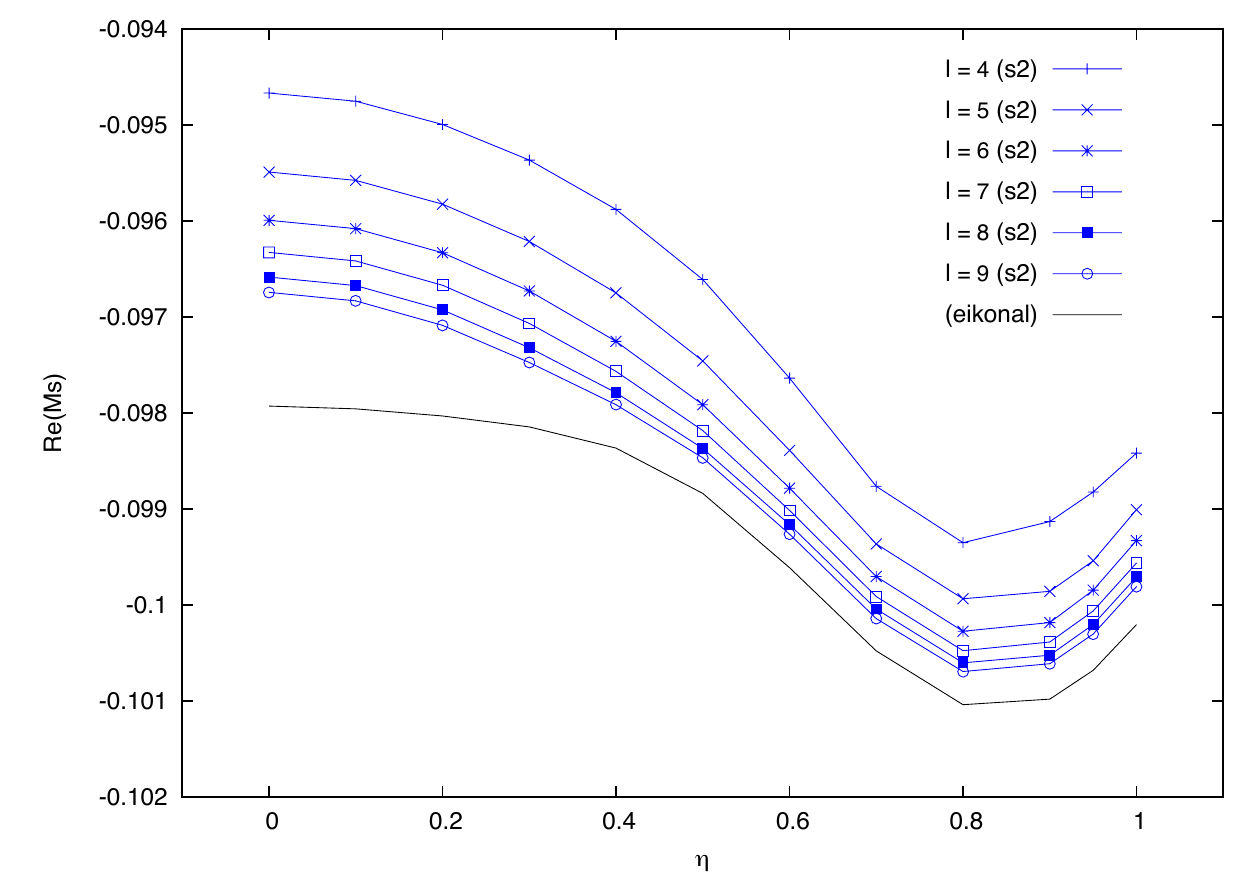} 
\includegraphics[height=2.4in]{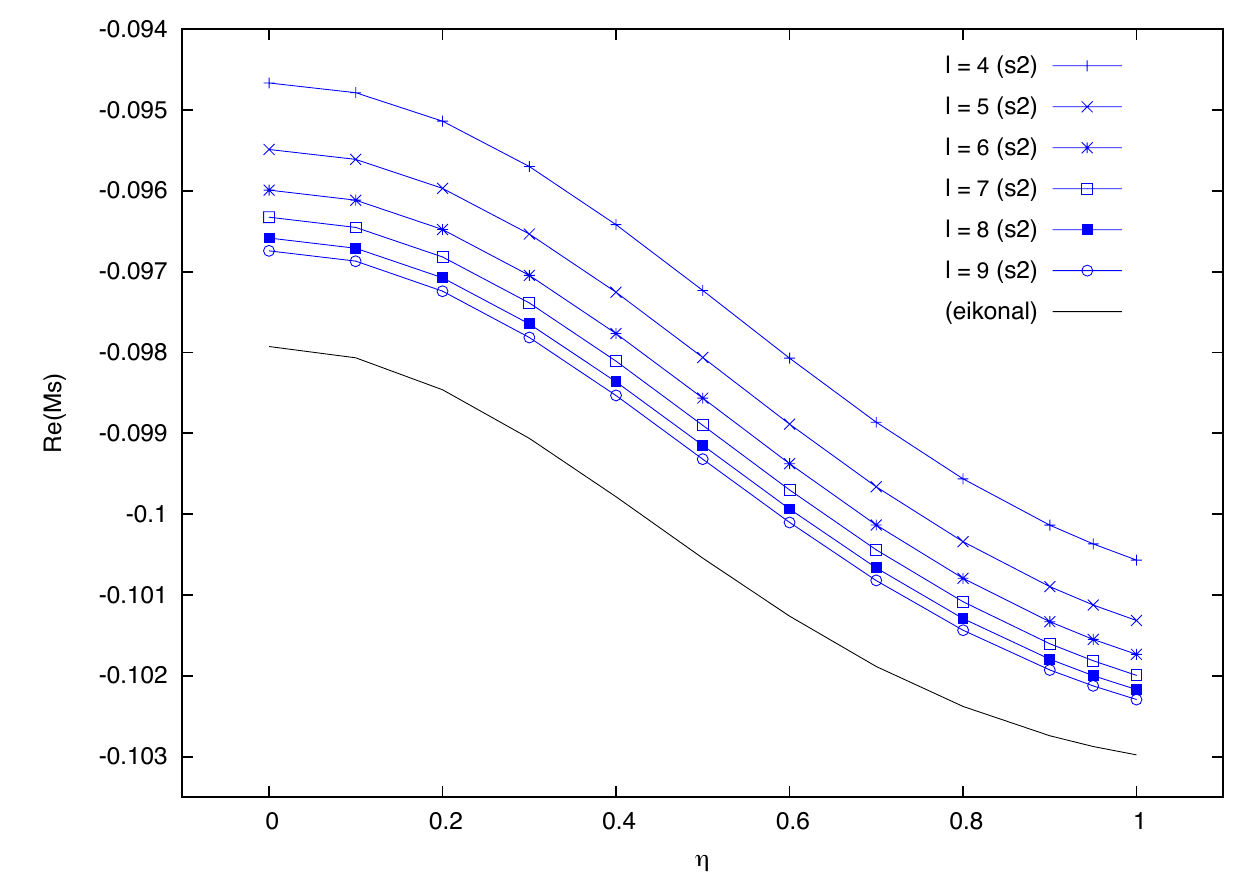}  
\caption{Decay rates for the frequency $s_2$ as a function of the deformation parameter $\eta$ and the corresponding result from the eikonal limit indicated here by the solid black line.
Left panel: Even parity sector, and the eikonal limit from the effective metric $\mathfrak{g}_1$. Right panel: Odd parity sector, and the eikonal limit from the effective metric $\mathfrak{g}_2$. As we see, for large values of $\ell$ the behavior in each parity sector is consistent with the predictions from the eikonal limit for the metrics $\mathfrak{g}_1$ and $\mathfrak{g}_2$, respectively.}
\label{Fig:Def2ReEta}
\end{figure}

In Figs.~\ref{Fig:Def1ReEta} and~\ref{Fig:Def2ReEta} we show the decay rates $\re(M s_1^{(\pm)})$ and $\re(M s_2^{(\pm)})$ for the deformed RN black holes versus the deformation parameter $\eta$, and compare them with the corresponding results from the eikonal limit, indicated by the solid black lines. We see that in both parity sectors $\re(M s_1)$ approaches the corresponding values computed in the eikonal limit from the spacetime metric $\mathfrak{g}_2 = {\bf g}$. In contrast to this, $\re(M s_2^{(+)})$ approaches the results for the effective metric $\mathfrak{g}_1$ while $\re(M s_2^{(-)})$ approaches the ones for the metric $\mathfrak{g}_2$.

 
\begin{figure}[h!]
\centering
\includegraphics[height=2.4in]{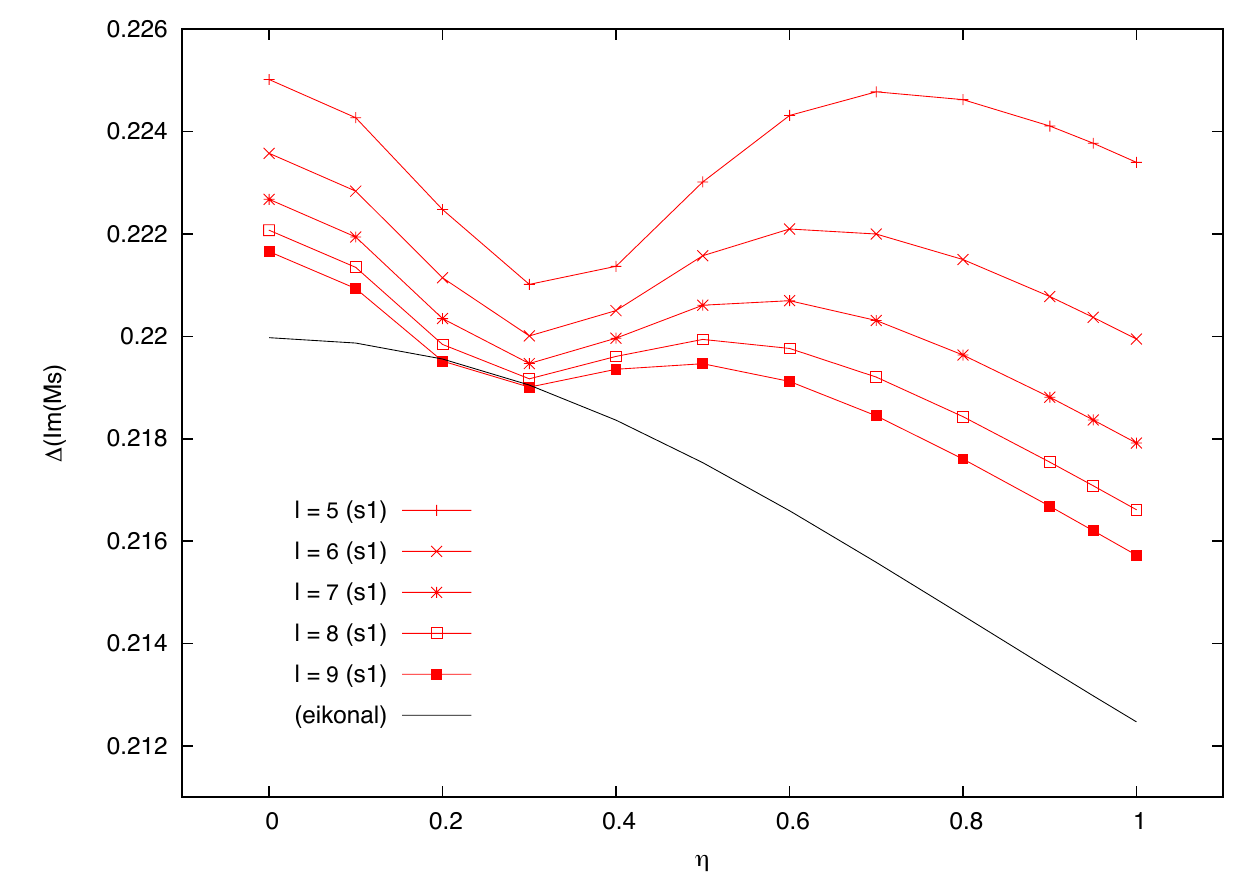} 
\includegraphics[height=2.4in]{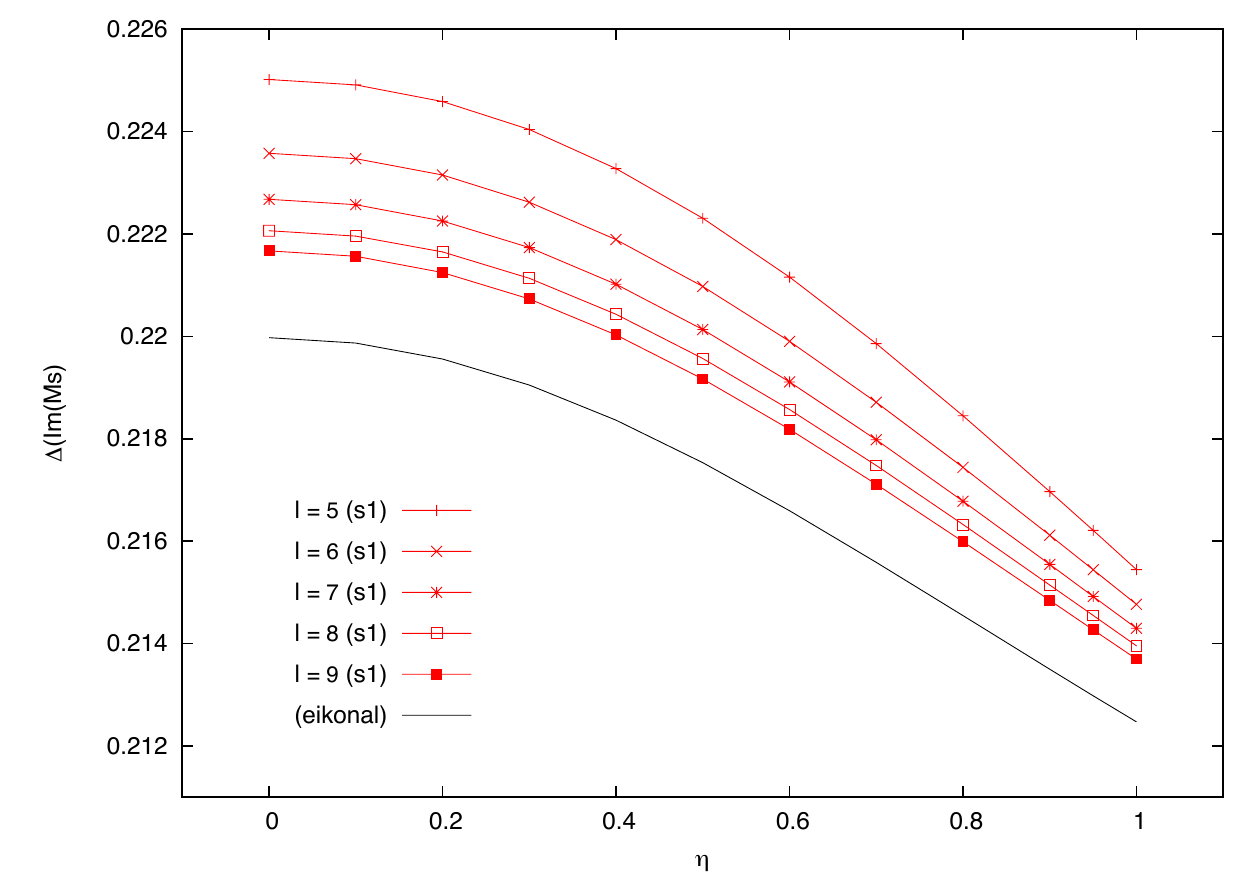}  
\caption{Difference of $\im(M s_1)$ between two successive values of $\ell$ vs $\eta$ and the corresponding result from the eikonal limit computed with the effective metric $\mathfrak{g}_2$ which are indicated by a solid black line.
Left panel: Even parity sector. Right panel: Odd parity sector.}
\label{Fig:Def1ImagEta}
\end{figure}

\begin{figure}[h!]
\centering
\includegraphics[height=2.4in]{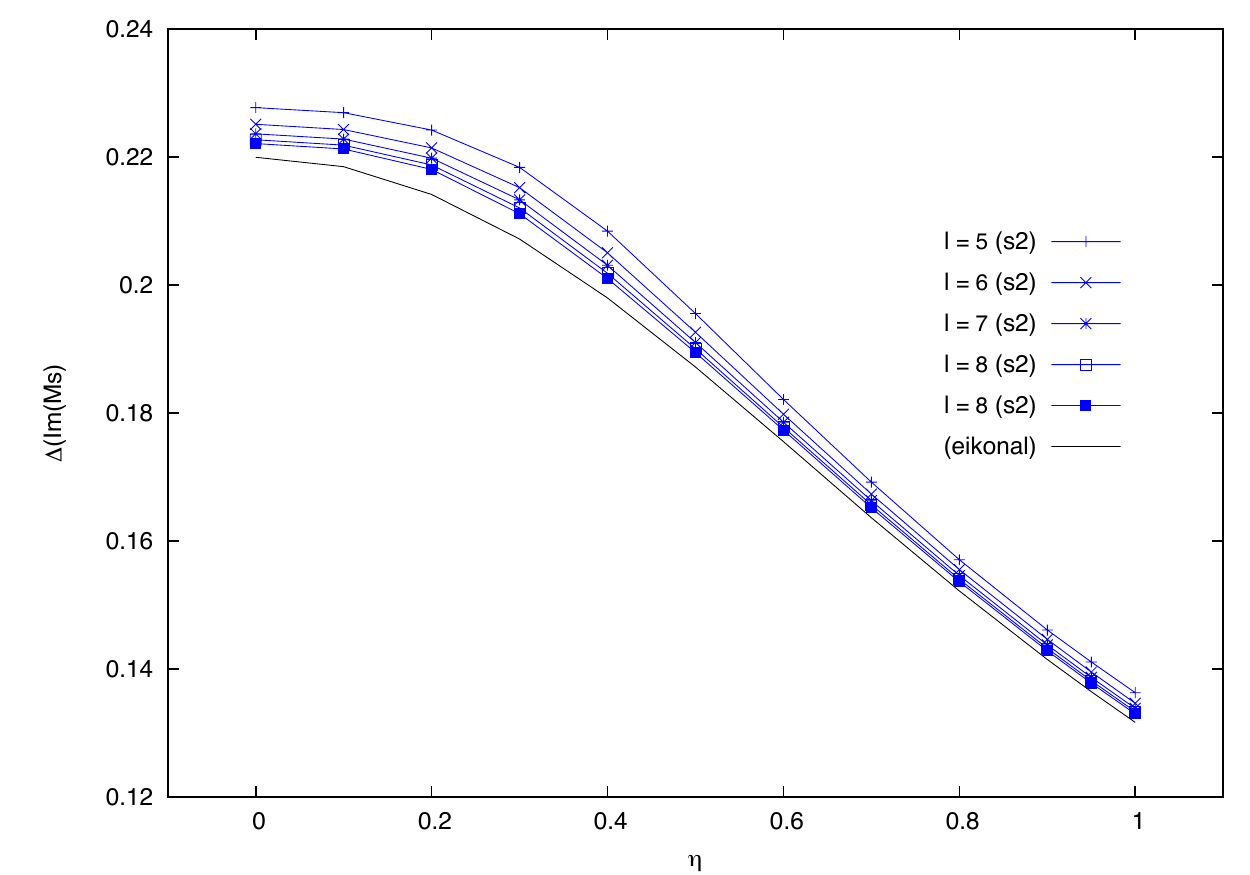} 
\includegraphics[height=2.4in]{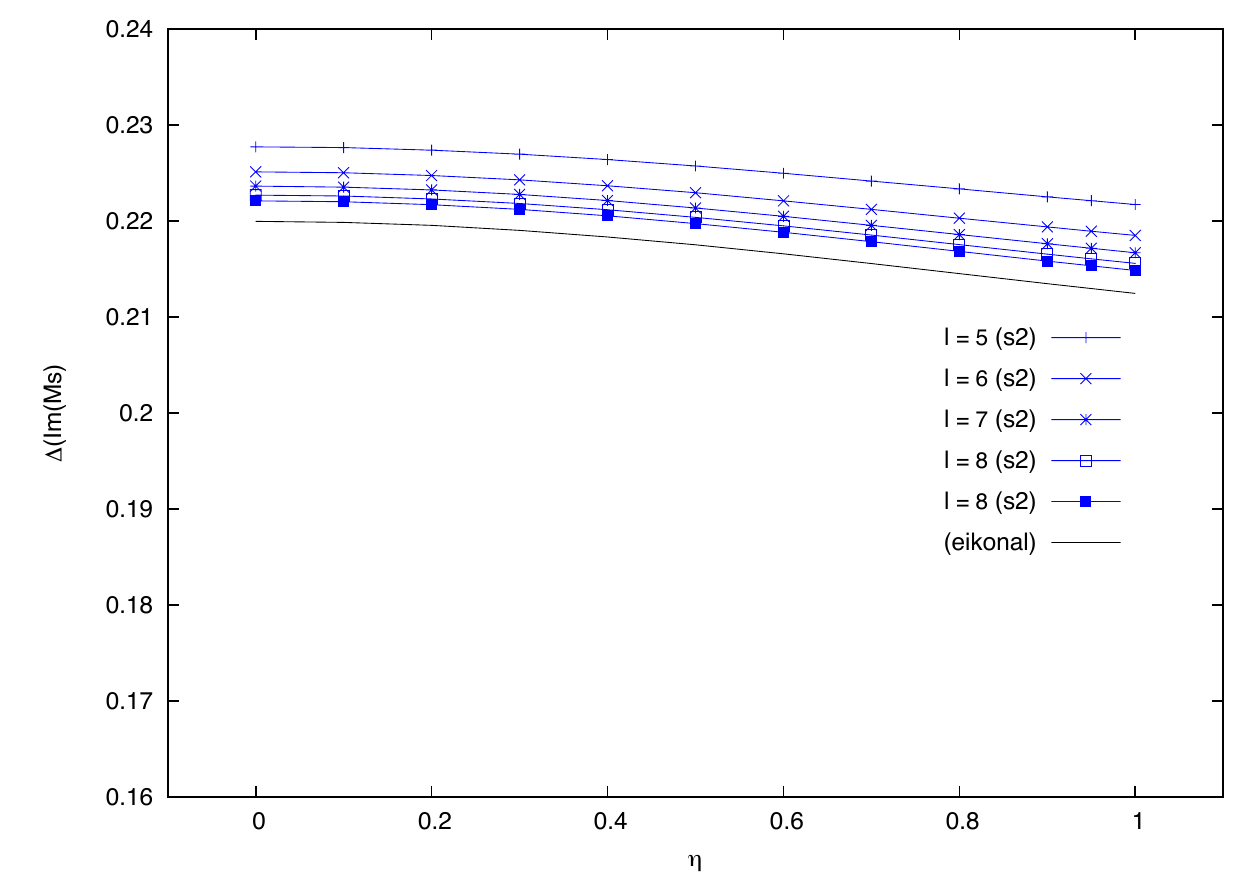}  
\caption{Difference of $\im(M s_2)$ between two successive values of $\ell$ vs $\eta$ and the corresponding result from the eikonal limit. Left panel: Even parity sector and eikonal limit computed using the spacetime metric $\mathfrak{g}_2$. Right panel: Odd parity sector and eikonal limit computed from the effective metric $\mathfrak{g}_1$.}
\label{Fig:Def2ImagEta}
\end{figure}

Finally, in Figs.~\ref{Fig:Def1ImagEta} and~\ref{Fig:Def2ImagEta} we show the difference in the imaginary parts 
of $s_1^{(\pm)}$ and $s_2^{(\pm)}$ between two successive values of $\ell$ as a function of $\eta$ for the deformed RN black hole. 
The corresponding result in the eikonal limit is indicated by the solid black line. As before, we see that in both parity 
sectors the results for $\im(M s_1)$ approach the value in the eikonal limit for the metric $\mathfrak{g}_2$ whereas $\im(M s_2^{(+)})$ 
approaches the value for $\mathfrak{g}_1$ and $\im(M s_2^{(-)})$ the value for $\mathfrak{g}_2$.

\section{Conclusions}
\label{Sec:Conclusions}

In the present work we studied linearized, dynamical perturbations of spherically symmetric, static and 
magnetically charged black holes arising in general relativity coupled to a nonlinear electromagnetic 
field. In recent years, such black holes have drawn quite a lot of interest due to their rich properties 
which can be quite different than the Reissner-Nordstr\"om (RN) black holes in the standard electrovacuum 
theory. Among these properties are the existence of regular black holes which do not contain a singularity 
at their center~\cite{eAaG98,eAaG99a,eAaG99b,eAaG00,eAaG05} and the existence of regular (soliton-like) 
solutions without horizons~\cite{kB01}.

Here, we performed the first consistent calculation for the quasi-normal (QN) frequencies of such black 
holes under the coupled gravitational and electromagnetic perturbations. Although there has been a lot of 
previous work regarding the calculation of QN frequencies of regular black holes, see for 
example~\cite{sFcH06,sFjC12,aFjL13,bTaAzSbA15,jLkLnY15,sF15,pXxA16}, the calculations so far apply only 
to \emph{test fields} (with spin $0$, $1$ or $2$) propagating on the fixed spacetime geometry associated 
with such holes. In contrast, in this article we considered linear perturbations of both the gravitational 
and electromagnetic fields in a self-consistent way, and thus our results refer directly to the vibrations 
of the black holes themselves, and not to those of the test fields.

Our calculations made use of the pulsation equations describing the coupled linear perturbations derived in~\cite{cMoS03}, where sufficient conditions for the linear stability of black 
holes in nonlinear electrodynamics (NED) were given. Due to the symmetry of the background, these equations 
constitute, in each parity sector, a family of radial wave equations for a vector-valued function describing 
the time evolution of two gauge-invariant linear combinations of the gravitational and electromagnetic 
perturbations. These two combinations are coupled to each other through the matrix-valued potential appearing 
in the radial wave equation. In the RN case of linear electromagnetism it is possible to decouple these 
perturbations, leading to two separate families of \emph{scalar} wave equations. Accordingly, there exist two 
families of QN spectra in each parity sector. Additionally, it turns out that in the RN case the modes in both 
parity sectors are \emph{isospectral}, that is, the odd and even potentials give rise to the same QN frequencies. 
This property is due to the existence of a special intertwining relation between the radial wave equations 
which remains valid in the Schwarzschild case~\cite{Chandrasekhar-Book}.

However, these special relations cease to exist in nonlinear electromagnetism and as a consequence, 
isospectrality is broken as we showed in this work. More specifically, we focused on a particular family of 
Lagrangians for the electromagnetic field which is described by a parameter $\eta$ which measures the strength 
of the nonlinearity of the field. When $\eta = 0$ the resulting spherically, symmetric static and purely magnetic 
solutions are the RN solutions and when $\eta > 0$ one obtains deformed RN black holes whose QN fundamental 
frequencies have a different spectrum in the even- and odd-parity sectors.

We performed our calculations using two different methods. The first method is based on a numerical Cauchy 
evolution of the pulsation equations derived in~\cite{cMoS03} and observing the signal measured by a static 
observer, while the second method consists in directly solving the ``eigenvalue'' problem associated with the 
QN modes using a generalization of the algorithm described in~\cite{eCmMoS15} to systems of equations. We first 
checked that both methods yielded the correct known frequencies in the RN case, and then computed some QN 
fundamental frequencies with both methods for the deformed RN solutions in NED, slowly increasing the value of 
the parameter $\eta$ and fixing the charge to mass ratio $q$. We found that both methods yield comparable results 
and both predict the parity splitting phenomena (see Table~\ref{Tab:freqRNDeformed}).

Finally, using the shooting method, we performed a more systematic study for the fundamental QN spectrum of the 
deformed black holes as a function of $\eta$ and the angular momentum number $\ell$. In particular, we analyzed 
the behavior of the frequencies for values of $\ell$ between $\ell = 2$ and $\ell = 9$ and compared our results 
with the eikonal (high-frequency) limit, in which the real and imaginary parts of the QN frequencies can be 
related to the properties of the unstable, circular null geodesics. An interesting observation is that whereas 
in standard electrodynamics the eikonal limit of electromagnetic radiation is always associated with null 
geodesics in the background spacetime $(M,{\bf g})$, in NED the light-rays in the geometric optics approximation 
follow null geodesics which are associated not with the spacetime gravitational metric ${\bf g}$ but rather with 
two effective metrics, $\mathfrak{g}_1$ and $\mathfrak{g}_2$ which are constructed from ${\bf g}$ and the 
electromagnetic field tensor ${\bf F}$ (see Refs.~\cite{yOgR02,fAfCeGoR15} and references therein). In the models 
we have considered in this article $\mathfrak{g}_2 = {\bf g}$, but $\mathfrak{g}_1 \neq \mathfrak{g}_2$. By comparing the properties of the unstable, circular null geodesics with respect to $\mathfrak{g}_1$ and $\mathfrak{g}_2$ to our numerically computed frequencies for the deformed RN black holes, we verified that as $\ell$ increases the latter approach the results corresponding to the eikonal limit. The fact that the two effective metrics $\mathfrak{g}_1 \neq \mathfrak{g}_2$ deviate more and more from each other as $\eta$ increases implies a breaking of isospectrality in the eikonal limit, and thus provides a partial explanation for the parity splitting phenomena found in this article.

A break of isospectrality in the QN spectrum was also found when studying linear perturbations of charged, dilaton black holes in Ref.~\cite{vFmPfP01}, based on the WKB approximation. However, in this case the break is expected because the presence of the scalar field describing the dilaton adds an additional perturbation equation in the even-parity sector, but not in the odd-parity one, so that it is a priori clear that the two parity sectors cannot be related to each other by an intertwining relation. In contrast to this, in the case we have studied here, the perturbation equations constitute a $2\times 2$ system of wave equations in both parity sector, like in the RN case.

It would be interesting to investigate in more detail for which matter models (or for which alternate theories of gravity) the parity splitting phenomena discussed in this article occurs.


\acknowledgments

We wish to thank Sharmanthie Fernando for comments on a previous version of our manuscript. This research was supported in part by CONACyT Grant No. 233137, by a CIC Grant to Universidad Michoacana, and by SNI-CONACyT. EC was partially supported by a PRODEP postdoctoral fellowship. JCD acknowledges support from Instituto de Ciencias F\'{\i}sicas, UNAM. CM thanks CONACYT-AEM Grant No.~248411 and Universidad de Guadalajara for academic and financial support.

\appendix
\section{Analysis of the solutions for the alternative model}
\label{App:SolutionProp}

In this appendix we provide a detailed analysis regarding the global behavior of the metric 
given in Eq.~(\ref{Eq:MetricSol}) which is characterized by the single function
\begin{equation}
N(r) = 1 - \frac{2m(r)}{r} = 1 - \frac{2M}{r} + \frac{Q_m^2}{2}\left[ 
 \frac{1}{r^2 + \eta^2} + \frac{\arctan\left( \frac{\eta}{r} \right)}{\eta r} \right],\qquad
r > 0,
\label{Eq:AlternativeBHBis}
\end{equation}
see Sec.~\ref{SubSec:Alternative}. In particular, our goal is to find out whether the corresponding metric describes a black hole or a soliton and whether or not it is regular at the center $r = 0$. In the limit $\eta\to 0$ the function $N$ reduces to the corresponding metric function of the RN metric. When $Q_ m = 0$, one obtains the Schwarzschild metric. In the following, we focus on the case where the three parameters $\eta$, $M$ and $Q_m$ are strictly positive.

\subsection{Global properties of the function $N$}

To analyze $N$, it is convenient to introduce the dimensionless quantities
$$
z := \frac{\eta}{r} > 0,\qquad
\beta := \frac{M}{\eta} > 0,\qquad
\delta := \frac{Q_m^2}{\eta^2} > 0,
$$
in terms of which $N(r) = F(z)$ with the function $F : [0,\infty)\to \Real$ defined by
\begin{equation}
F(z) = 1 - 2\beta z + \frac{\delta}{2}\left[ \frac{z^2}{z^2 + 1} + z\arctan(z) \right],\qquad
z\geq 0.
\label{Eq:FDef}
\end{equation}
For small $z$ we have
$$
F(z) = 1 - 2\beta z + \delta z^2 + {\cal O}(z^4),
$$
which reflects the RN asymptotics of the solution. For large $z$ we have
$$
\lim\limits_{z\to \infty} \frac{F(z)}{z} = \delta\frac{\pi}{4} - 2\beta.
$$
Furthermore, the first and second derivatives of $F$ are given by
\begin{eqnarray*}
F'(z) &=& -2\beta + \frac{\delta}{2}\left[ \frac{3z + z^3}{(z^2 + 1)^2} + \arctan(z) \right],\\
F''(z) &=& 2\delta\frac{1 - z^2}{(z^2 + 1)^3},
\end{eqnarray*}
and hence $F$ is convex for $0 < z < 1$ and concave for $z > 1$. At the inflection point $z = 1$ one has
$$
F(1) = 1 - 2\beta + \frac{\delta}{2}\left[ \frac{1}{2} + \frac{\pi}{4} \right],\qquad
F'(1) = -2\beta + \frac{\delta}{2}\left[ 1 + \frac{\pi}{4} \right].
$$
Based on these remarks, we can prove the following:

\begin{lemma}
\label{Lem:BehaviorF}
The function $F: [0,\infty)\to \Real$ defined in Eq.~(\ref{Eq:FDef}) has the following behavior:
\begin{enumerate}
\item[(a)] For $\frac{\beta}{\delta}\geq \frac{1}{4}\left( 1 + \frac{\pi}{4} \right)$ the function $F$ is strictly monotonously decreasing from $1$ to $-\infty$, and thus it has precisely one zero.
\item[(b)] When $\frac{\pi}{8} < \frac{\beta}{\delta} < \frac{1}{4}\left( 1 + \frac{\pi}{4} \right)$ the function $F$ has a unique local minimum in the region $0 < z < 1$, a unique local maximum in the region $z > 1$, and $F(z)\to -\infty$ as $z\to \infty$. In this case $F$ can have either one, two or three zeros, depending on the values of $F$ at the local extrema.
\item[(c)] For $\frac{\beta}{\delta} < \frac{\pi}{8}$ the function $F$ has a unique global minimum in the region $0 < z < 1$. Furthermore, $F$ is strictly increasing for $z > 1$ with $F(z)\to +\infty$ as $z\to \infty$. Consequently, $F$ has either no zeros, one degenerate zero or precisely two zeros, depending on whether the value of $F$ at the minimum is positive, zero or negative.
\item[(d)] For $\frac{\beta}{\delta} = \frac{\pi}{8}$ one has the same properties as in the previous case except that $F(z) \to 1$ as $z\to \infty$.
\end{enumerate}
\end{lemma}

\proof In order to prove (a) we first note that $F(0) = 1$, $F'(0) = -2\beta$, and
$$
\lim\limits_{z\to \infty} \frac{F(z)}{z} = \lim\limits_{z\to \infty} F'(z)
 = \delta\frac{\pi}{4} - 2\beta\leq -\frac{\delta}{2}\left( 1 - \frac{\pi}{4} \right) < 0,
$$
so that $F(z)\to -\infty$ as $z\to \infty$. Furthermore, since
$$
F'(1) = -2\beta + \frac{\delta}{2}\left[ 1 + \frac{\pi}{4} \right] \leq 0,
$$
and because $F$ is convex for $0 < z < 1$ and concave for $z > 1$, it follows that $F'(z) < 0$ for all $z\geq 0$ except possibly at $z = 1$, and hence $F$ is strictly monotonously decreasing.

To prove (b) we note that we still have $\lim\limits_{z\to \infty} F'(z) < 0$ and $F'(0) = -2\beta < 0$, but now the hypothesis implies that $F'(1) > 0$. Therefore, because $F$ is convex for $0 < z < 1$ there is a unique local minimum in this region, and because $F$ is concave for $z > 1$ there is a unique local maximum in that region.

Finally, to prove (c) and (d) we note that in these cases $F'(0) = -2\beta < 0$ and $F'(1)\geq \delta(1 - \pi/4)/2 > 0$, so there is again a unique local minimum in the region $0 < z < 1$. However, in contrast to the previous cases, $F'(z) \to \delta\pi/4 - 2\beta\geq 0$ as $z\to \infty$ and because $F$ is concave for $z > 1$ it follows that $F'(z) > 0$ for all $z > 1$. This implies that the minimum in the region $0 < z < 1$ is global. Furthermore, it follows that $F(z)\to +\infty$ as $z\to \infty$ when $\beta/\delta < \pi/8$ while for $\beta/\delta = \pi/8$ one finds the asymptotic expansion
$$
F(z) = 1 - \frac{\delta}{3z^2} + {\cal O}\left( \frac{1}{z^4} \right).
$$
\qed

The minimum of $F$ in cases (b)--(d) is located at $z = z_{min}$, where $0 < z_{min} < 1$ is uniquely determined by the equation
$$
\frac{1}{4}\left[ \frac{3z + z^3}{(z^2+1)^2} + \arctan(z) \right]_{z=z_{min}} = \frac{\beta}{\delta},
$$
and
$$
F(z_{min}) = 1 - \delta\left. \frac{z^2}{(z^2 + 1)^2} \right|_{z=z_{min}}.
$$
Hence, we see that while the location of the minimum depends only on the ration $\beta/\delta$, the fact of whether or not this minimum is positive depends only on the value of $\delta$.

\subsection{Global behavior of the metric}

In order to give a geometric interpretation of the results formulated in the previous lemma, we first note that the function $F$ is equal to minus the norm of the asymptotically timelike Killing vector field. Furthermore, $F$ determines the geometry of the two-manifolds $(\tilde{M},\gtiltens)$ which are orthogonal to the invariant two-spheres. For instance, the Gauss curvature of $(\tilde{M},\gtiltens)$ is
\begin{equation}
\tilde{\kappa} = -\frac{1}{2} \frac{d^2 N}{dr^2} 
 = \frac{2\delta}{\eta^2} z^3\left\{ \frac{\beta}{\delta}  
 - \frac{1}{4}\left[ \arctan(z) + z \frac{5 + 2z^2 + z^4}{(1+z^2)^3} \right] \right\},
\end{equation}
and using the expansion $\arctan(z) = \pi/2 - 1/z + 1/(3z^3) - 1/(5z^5) + \ldots$ one finds
\begin{equation}
\tilde{\kappa} = \frac{2\delta}{\eta^2}\left\{ \left( \frac{\beta}{\delta} - \frac{\pi}{8} \right) z^3 
 + \frac{1}{6} + {\cal O}\left( \frac{1}{z^2} \right) \right\},
\end{equation}
which diverges as one approaches the origin $r\to 0$ (that is, $z\to \infty$), unless $\beta/\delta = \pi/8$.

Therefore, Lemma~\ref{Lem:BehaviorF} yields the following results: Cases (a) and (b) correspond to singular black hole solutions with an event horizon and possibly also inner horizons. Case (c) corresponds to either a singular black hole or to a naked singularity, depending on whether the minimum of $F$ is negative or positive. Finally, case (d) corresponds either to a regular black hole or to a globally regular solution, depending on whether the minimum of $F$ is negative or positive.

As an application, consider the parametrization of the solutions in terms of the dimensionless parameters $\overline{\eta}$ and $q$ introduced at the beginning of Sec.~\ref{Sec:Results}, in terms of which
$$
\beta = \frac{M}{\eta} = \frac{\overline{M}}{\overline{\eta}},\qquad
\delta = \frac{Q_m^2}{\eta^2} = q^2\frac{\overline{M}^2}{\overline{\eta}^2},\qquad
\overline{M} = \frac{1}{1 + \sqrt{1 - E(\overline{\eta}) q^2}}.
$$
Therefore,
$$
\frac{\beta}{\delta} = \frac{\overline{\eta}}{q^2}\left( 1 + \sqrt{1 - E(\overline{\eta}) q^2} \right),
$$
and depending on the values for $\overline{\eta}$ and $q$ we see that any of the cases (a)--(d) in Lemma~\ref{Lem:BehaviorF} can occur. For example, for the choice $q = 0.8$ considered in Sec.~\ref{Sec:Results} the value for $\beta/\delta$ varies between $0$ and about $2.7613$ as $\overline{\eta}$ sweeps through the interval $[0,1]$. At the horizon,
$$
z = z_H = \frac{\eta}{r_H} = \overline{\eta},
$$
and we can check by explicit calculation that $F(\overline{\eta}) = 0$ and that
$$
\overline{\eta} F'(\overline{\eta}) = -1 + \frac{q^2\overline{M}^2}{(1 + \overline{\eta}^2)^2}.
$$
Therefore, by choosing $q^2 < 1$ and $0 < \overline{\eta} < 1$ we can guarantee (since $\overline{M} \leq 1$) that $F'(\overline{\eta}) < 0$, and hence $z = \overline{\eta} < 1$ is indeed the smallest zero of $F$, corresponding to the location of the event horizon.

Finally, we comment on the fulfillment of the conditions~(\ref{Eq:StabilityCond}) which guarantee the linear stability of the black holes. Since ${\cal L} > 0$ and ${\cal L}_y > 0$ are automatically satisfied for the alternative model, these conditions are equivalent to
$$
\kappa > 0\hbox{ and } N\kappa \leq 3,
$$
for all $r \geq r_H$. Since $N(r) = F(z)\in (0,1]$ for all $0\leq z < z_H$, and since $\kappa = (1 - 2z^2)/(1 + z^2)\leq 1$, the second condition is automatically satisfied. The first condition is also satisfied provided that
$$
0\leq \overline{\eta} < \frac{1}{\sqrt{2}}.
$$
On the other hand, the deformed RN black hole discussed in Sec.~\ref{SubSec:Alternative} are linearly unstable when $\overline{\eta} > 1/\sqrt{2}$ since in this case $\kappa < 0$ near the event horizon.

\bibliographystyle{unsrt}
\bibliography{refs_pert}

\end{document}